\RequirePackage{silence}
\documentclass[%
 reprint,
 superscriptaddress,
nofootinbib,
 amsmath,amssymb,
 aps,
 pra,
floatfix,
longbibliography 
]{revtex4-2}

\usepackage{graphicx}
\usepackage{dcolumn}
\usepackage{bm}
\usepackage[breaklinks=true]{hyperref}
\hypersetup{
    colorlinks,
    linkcolor={red!80!black},
    citecolor={green!80!black},
    urlcolor={blue!80!black}
}

\usepackage{enumerate}
\usepackage[inline]{enumitem}
\usepackage{tikz}
\usetikzlibrary{shapes, arrows.meta}
\usetikzlibrary{quantikz2}

\usepackage{blindtext}
\usepackage{amsthm}
\usepackage{mathtools}
\DeclarePairedDelimiter{\ceil}{\lceil}{\rceil}

\DeclareMathOperator*{\argmin}{arg\,min}
\usepackage{braket}
\theoremstyle{plain}
\newtheorem{theorem}{Theorem}
\newtheorem{corollary}{Corollary}
\newtheorem{lemma}{Lemma}
\newtheorem{definition}{Definition}
\theoremstyle{definition}
\newtheorem{remark}{Remark}

\usepackage{nicefrac}
\usepackage{cleveref}
\Crefname{section}{Sec.}{Secs.}
\Crefname{theorem}{Thm.}{Thms.}
\Crefname{lemma}{Lem.}{Lems.}
\Crefname{corollary}{Cor.}{Cors.}
\Crefname{definition}{Def.}{Defs.}
\Crefname{remark}{Rem.}{Rems.}
\Crefname{figure}{Fig.}{Figs.}
\Crefname{equation}{Eq.}{Eqs.}

\usepackage[caption=false]{subfig}

\DeclareMathOperator{\polylog}{polylog}
\tikzset{%
    uctrl/.style={draw, circle, minimum size=0.285pc, append after command={ \pgfextra { \fill  (0,0)-- (225:2pt) arc (225:405:2pt) -- cycle; } } }
}
\DeclareExpandableDocumentCommand{\uctrl}{O{}{m}}{|[uctrl,#1]| {#2} \qw}

\tikzset{%
    pctrl/.style={draw, circle, minimum size=0pc, append after command={ \pgfextra { \path[draw=black,fill=white]   (-0.075,0.055)-- (0.075,0.055) -- (0,-0.095) -- (-0.075,0.055) -- cycle; } } }
}
\DeclareExpandableDocumentCommand{\pctrl}{O{}{m}}{|[pctrl,#1]| {#2} \qw}

\tikzset{%
    zctrl/.style={draw, circle, minimum size=0.285pc, fill=white, append after command={ \pgfextra { \path[draw=black,fill=black]  (0:0.2pt) arc (0:360:0.2pt) -- cycle; } } }
}
\DeclareExpandableDocumentCommand{\zctrl}{O{}{m}}{|[zctrl,#1]| {#2} \qw}

\newsavebox{\boxQ}
\AtBeginDocument{
    \newwrite\bibnotes
    \def\bibnotesext{Notes.bib}
    \immediate\openout\bibnotes=\jobname\bibnotesext
    \immediate\write\bibnotes{@CONTROL{REVTEX42Control}}
    \immediate\write\bibnotes{@CONTROL{
    apsrev42Control,author="48",editor="1",pages="1",title="0",year="1"}}
     \if@filesw
     \immediate\write\@auxout{\string\citation{apsrev42Control}}
    \fi
}

\usetikzlibrary{chains}

\makeatletter
\DeclareRobustCommand{\rvdots}{%
  \vbox{
    \baselineskip4\p@\lineskiplimit\z@
    \kern-\p@
    \hbox{.}\hbox{.}\hbox{.}
  }}
\makeatother

\makeatletter
\DeclareRobustCommand{\rvdotstwo}{%
  \vbox{
    \vspace{3pt}
    \baselineskip3\p@\lineskiplimit\z@
    \kern-\p@
    \hbox{.}\hbox{.}\hbox{.}
    \vspace{3pt}
  }}
\makeatother

\makeatletter
\DeclareRobustCommand
  \rddots{\mathinner{\mkern1mu\raise7\p@
    \vbox{\hbox{.}}\mkern2mu
    \raise4\p@\hbox{.}\mkern2mu\raise\p@\hbox{.}\mkern1mu}}
\makeatother

\makeatletter
\DeclareRobustCommand
  \lddots{\mathinner{
  \mkern2mu\raise\p@\hbox{.}
  \mkern2mu\raise4\p@\hbox{.}
  \mkern1mu\raise7\p@\vbox{\hbox{.}}\mkern1mu}}
\makeatother

\makeatletter
\def\thickhline{%
  \noalign{\ifnum0=`}\fi\hrule \@height \thickarrayrulewidth \futurelet
   \reserved@a\@xthickhline}
\def\@xthickhline{\ifx\reserved@a\thickhline
               \vskip\doublerulesep
               \vskip-\thickarrayrulewidth
             \fi
      \ifnum0=`{\fi}}
\makeatother

\newlength{\thickarrayrulewidth}
\makeatletter
\renewcommand{\qwbundle}[2][]{
  \pgfkeys{/quantikz/gates/.cd,style=,Strike Width=0.08cm,Strike Height=0.12cm,#1}%
  \pgfkeysgetvalue{/quantikz/gates/style}{\qz@style}%
  \pgfkeysgetvalue{/quantikz/gates/Strike Width}{\qz@sw}%
  \pgfkeysgetvalue{/quantikz/gates/Strike Height}{\qz@sh}%
  \expanded{%
    \noexpand\arrow[strike arrow={\qz@sw}{\qz@sh}{\unexpanded{#2}},\qz@style,phantom]{l}%
  }%
}
\makeatother

\makeatletter
\DeclareRobustCommand{\pbar}{\mathord{%
  \text{$\m@th\mkern-2mu\raisebox{-1.5ex}[0pt][0pt]{$\mathchar'26$}\mkern-7mu p$}%
}}
\makeatother

\makeatletter
\DeclareRobustCommand{\Bbar}{\mathord{%
  \text{$\m@th\mkern-2mu\raisebox{-0.85ex}[0pt][0pt]{$\mathchar'26$}\mkern-9mu B$}%
}}
\makeatother

\makeatletter
\DeclareRobustCommand{\Bbarc}{\mathord{%
  \text{$\m@th\mkern-2mu\raisebox{-0.85ex}[0pt][0pt]{$\mathchar'26$}\mkern-8mu \mathcal{B}$}%
}}
\makeatother

\makeatletter
\DeclareRobustCommand{\Lbarc}{\mathord{%
  \text{$\m@th\mkern-2mu\raisebox{-0.6ex}[0pt][0pt]{$\mathchar'26$}\mkern-9mu \mathcal{L}$}%
}}
\makeatother

\begin{document}

\title{Exponential Quantum Speedup for Simulation-Based Optimization Applications}

\author{Jonas Stein}
 \email{jonas.stein@ifi.lmu.de}
 \affiliation{Institute for Informatics, LMU Munich}
 \affiliation{Aqarios GmbH, Munich}
\author{Lukas Müller}
 \affiliation{BMW Group, Munich}
\author{Leonhard Hölscher}
 \affiliation{BMW Group, Munich}
\author{Georgios Chnitidis}
 \affiliation{BMW Group, Munich}
\author{Jezer Jojo}
 \affiliation{Indian Institute of Science Education and Research Pune}
\author{Afrah Farea}
 \affiliation{Istanbul Technical University}
\author{Mustafa Serdar Çelebi}
 \affiliation{Istanbul Technical University}
\author{David Bucher}
 \affiliation{Aqarios GmbH, Munich}
 \author{Jonathan Wulf}
 \affiliation{Institute for Informatics, LMU Munich}
 \author{David Fischer}
 \affiliation{Institute for Informatics, LMU Munich}
\author{Philipp Altmann}
 \affiliation{Institute for Informatics, LMU Munich}
\author{Claudia Linnhoff-Popien}
 \affiliation{Institute for Informatics, LMU Munich}
\author{Sebastian Feld}
 \affiliation{Quantum \& Computer Engineering, Delft University of Technology}
\date{\today}

\begin{abstract}
The simulation of many industrially relevant physical processes can be executed up to exponentially faster using quantum algorithms. However, this speedup can only be leveraged if the data input and output of the simulation can be implemented efficiently. While we show that recent advancements for optimal state preparation can effectively solve the problem of data input at a moderate cost of ancillary qubits in many cases, the output problem can provably not be solved efficiently in general. By acknowledging that many simulation problems arise only as a subproblem of a larger optimization problem in many practical applications however, we identify and define a class of practically relevant problems that does not suffer from the output problem: Quantum Simulation-based Optimization (QuSO). QuSO represents optimization problems whose objective function and/or constraints depend on summary statistic information on the result of a simulation, i.e., information that can be efficiently extracted from a quantum state vector. In this article, we focus on the LinQuSO subclass of QuSO, which is characterized by the linearity of the simulation problem, i.e., the simulation problem can be formulated as a system of linear equations. By cleverly combining the quantum singular value transformation (QSVT) with the quantum approximate optimization algorithm (QAOA), we prove that a large subgroup of LinQuSO problems can be solved with up to exponential quantum speedups with regards to their simulation component. Finally, we present two practically relevant use cases that fall within this subgroup of QuSO problems. 
\end{abstract}

\maketitle


\section{Introduction}
\label{sec:introduction}
Initially motivated by Richard Feynman~\cite{Feynman1982}, one of the main quantum applications with provable exponential speedups is the simulation of physical and chemical systems. While this idea was historically targeted at the simulation of quantum mechanical systems ~\cite{Lloyd96}, the discovery of a quantum algorithm for solving systems of linear equations, that also provides an exponential speedup in terms of the number of unknowns~\cite{HHL,Ambainis12,Berry15,Childs17,gilyen2018quantum,SSO19,Lin2020optimalpolynomial,posDefQLSP}, extended this notion to the simulation of many classical physical systems. Since then, significant quantum speedups have been proven for many (potentially) practically relevant problems~\cite{Cao_2013,PhysRevA.99.012323,Linden2022,PhysRevX.13.041041}, based on techniques to formulate differential equations into an efficiently solvable format for quantum algorithms~\cite{Childs2021highprecision,jin2022quantumsimulationpartialdifferential}. Through linearization~\cite{leyton2008quantumalgorithmsolvenonlinear,lloyd2020quantumalgorithmnonlineardifferential,doi:10.1073/pnas.2026805118} or linear representations~\cite{PhysRevResearch.2.043102,10.1063/5.0056974,JIN2024103457}, even many non-linear differential equations have been shown to allow for exponential quantum speedups.

In the simulation of linear classical physical systems the biggest concern has historically been the data input to and the output from the quantum algorithm (cf., e.g., Ref.~\cite{HHL}).

Concerning the input problem, most literature either assumes the existence of quantum oracles providing access to the inputs or even QRAM~\cite{Kerenidis2017} (for which efficient hardware realization has shown to be very difficult). While the input of specific, mathematically highly structured matrices is starting to be explored (cf. \cite{guseynov2024explicitgateconstructionblockencoding,doi:10.1137/22M1484298}), generally applicable approaches have not yet been established. As our first key contribution of this article, we show how recent results on provably optimal quantum state preparation \cite{10044235,Yuan2023optimalcontrolled} can be used to efficiently prepare sparse matrices by combining them with the encoding approach from Ref.~\cite{gilyen2018quantum}.

Addressing the output problem, it is important to acknowledge that any form of quantum state tomography necessary to extract all amplitudes of a given $n$-qubit quantum state vector takes $\mathcal{O}(2^n)$ many steps~\cite{paris2004quantum}, which destroys any exponential quantum speedup achieved during the computation of this state. However, it is possible to efficiently extract some information from a quantum state using quantum algorithms like quantum phase estimation or the Hadamard test. As our second core contribution, we compile a comprehensive framework of quantum algorithms that can efficiently extract information (often called summary statistic information, cf.~\cite{HHL}) from a state vector.

By acknowledging that real world academic and industrial simulation problems often occur in context of a larger optimization problem, one can recognize that the relevant information about the result of the simulation problem can in fact be efficiently extracted by a quantum algorithm for many practically relevant use cases. Concrete examples entail so-called simulation-based optimization problems, i.e., problems about finding input parameters to a simulation model that optimize a given objective\footnote{Note that a large body of research in simulation-based optimization investigates simulations of stochastic functions, which we do not -- we exclusively consider deterministic simulations. More precisely, we only consider parametric (i.e., static) optimization, i.e., we only consider the system of interest's properties in one specific real world configuration (cf. \cite{Gosavi2015,Amaran2014,Trigueiro2019,TEKIN2004}).}~\cite{WANG2013}, which can be found in many areas of application and research such as pharmaceutical development~\cite{myers2016response} or aircraft design~\cite{balabanov1996topology}. We later formalize the class of simulation-based optimization problems that do not suffer from the output problem via a corresponding definition of Quantum Simulation-based Optimization (QuSO). The central problem inherent to simulation-based optimization is that the computation of the objective value for every probed solution is very costly, such that often, only a very small subset of possible solutions can be explored (see e.g. Refs.~\cite{Carson1997,April2003,WANG2013}).

As our main contribution in this article, we show that exactly this connection of simulation and optimization components makes simulation-based optimization problems extremely well suited to quantum computing. We show that by cleverly combining the state-of-the-art quantum algorithms for optimization (QAOA) and solving systems of linear equations (QSVT), the output problem can be bypassed and the search space can be efficiently explored. Further, we provide an extensive complexity analysis and exemplify its practical application for two industrially relevant use cases: The unit commitment problem focused on optimal power flow and a basic form of topology optimization.

\textbf{Notation}: For arbitrary $n\in\mathbb{N}$, we use the notation $\left[n\right]\coloneq \left\lbrace 1, ..., n\right\rbrace$. We write $\mathbb{I}$ for an identity matrix of size that is clear by the context. In quantum circuits, we denote a \mbox{(multi-)}controlled $Z$-gate with connected $\bullet$ symbols (unless they are connected to any other gates, then this symbol denotes the usual control-operator). This is well-defined, as the position of the $Z$-gate in a \mbox{(multi-)}controlled $Z$-gate does not change the operator. If the controlled gate should actuate on the $\ket{0}$ state, the $\circ$ symbol is used instead. We use $||\cdot||_2$ to denote the spectral norm, i.e., the largest singular value of the given matrix.

\section{Preliminaries}
\label{sec:preliminaries}
This section provides a formal definition of QuSO, basics on quantum optimization with the QAOA and established (partly reformulated) quantum algorithms for data input, data processing, and data output relevant for solving QuSO problems.

\subsection{Quantum Simulation-based Optimization}
In the following, we define the set of QuSO problems as a subset of MINLP, the most general group of practically relevant optimization problems. Note that we only consider minimization problems wlog.

\begin{definition}[MINLP]\label{def:MINLP}
A Mixed-Integer Nonlinear Programming (MINLP) problem is an optimization problem of the form 
\begin{align*}
    &\underset{{x}}{\textnormal{minimize}}  &&  f(x)\\
    &\textnormal{subject to} && c_j(x)\leq  0,\;\forall j\in \left[K\right],\\
    & && x_i\in \left[l_i,u_i\right]\subset\mathbb{R}, \\
    & && x_i\in \mathbb{Z},\;\forall i\in I\subseteq \left[N\right],
\end{align*}
where $f:\mathbb{R}^N \rightarrow \mathbb{R}$ and $c:\mathbb{R}^N \rightarrow \mathbb{R}^K$ are continuous functions.
\end{definition}
As we define QuSO based on the concept of summary statistic information, we now formally state what is generally understood under this term.

\begin{definition}[Summary Statistic Information]\label{def:SSI}
    Given an oracle $O$ that prepares an $n$-qubit quantum state $\ket{\psi}=O\ket{0}^{\otimes n}$, we define summary statistic information about $\ket{\psi}$ as the output of a quantum algorithm that yields a basis-encoded binary string that depends on $\ket{\psi}$ given access to $O$.
\end{definition}

\begin{definition}[Quantum simulation-based optimization]\label{def:QuSO}
We define a quantum simulation-based optimization (QuSO) problem as a MINLP problem whose objective function and/or constraints depend on the summary statistic result of a simulation problem, i.e.,
\begin{align*}
    &\underset{{x}}{\textnormal{minimize}}  &&  f(x, u(s(x)))\\
    &\textnormal{subject to} && c_j(x, u(s(x)))\leq  0,\;\forall j\in \left[K\right],\\
    & && x_i\in \left[l_i,u_i\right]\subset\mathbb{R}, \\
    & && x_i\in \mathbb{Z},\;\forall i\in I\subseteq \left[N\right],
\end{align*}
where $f:\mathbb{R}^N \times \left\lbrace0,1\right\rbrace^m \rightarrow \mathbb{R}$ and $c:\mathbb{R}^N \times \left\lbrace0,1\right\rbrace^m \rightarrow \mathbb{R}^K$ are continuous functions, and $s:\mathbb{R}^N \rightarrow \mathbb{R}^M$ represents a simulation problem of which summary statistic information can be efficiently extracted via the function $u: \mathbb{R}^M \rightarrow \left\lbrace0,1\right\rbrace^m$.
\end{definition}

This class of optimization problems is designed to entail all simulation-based optimization problems that could potentially allow for a quantum speedup for their simulation components. Apart from our area of focus in this paper---linear simulation problems---QuSO also contains important problems involving non-linear simulations, e.g., based on the Navier-Stokes equations~\cite{Gaitan2020}. Note that QuSO does not differentiate between ''quantum problems``, like finding the ground state of an electronic structure, or ''classical problems`` like the simulation of structural mechanics, QuSO merely specifies that the information about the result of the simulation problem required by the optimization problem can be extracted  efficiently. In this context, \emph{efficiently} means that the computational complexity of extracting the summary statistic information does not outweigh a potential quantum speedup gained through a faster simulation.

\begin{definition}[Linear Quantum simulation-based optimization]\label{def:LinQuSO}
We define a linear quantum simulation-based optimization (LinQuSO) problem as a QuSO problem for which the underlying simulation problem takes the form of a system of linear equations (SLE) for all possible solutions.
\end{definition}

In this paper, we provide a framework to construct a quantum algorithm to solve LinQuSO problems of many common forms, allowing for an exponential speedup of the simulation component if all of the following conditions are satisfied.
\begin{enumerate}
    \item The SLE is sparse and well-conditioned.
    \item The dependence of the SLE on the decision variables allows for an efficient input to a quantum linear system solver. (For details see \Cref{subsec:meth:data-input})
    \item The extraction of summary statistic information from the result of the SLE can be done as efficiently as the simulation. (For details see \Cref{subsec:meth:data-readout})
\end{enumerate}



\subsection{Quantum Optimization}\label{subsec:quantopt}
As an approximated form of the Quantum Adiabatic Algorithm (QAA)~\cite{farhi2000quantum}, the QAOA utilizes the Adiabatic Theorem~\cite{Born1928} to approximate the solutions of unconstrained\footnote{To incorporate constraints, the standard approach penalty terms can be introduced (see, e.g., Ref.~\cite{Lucas2014}), or more sophisticated approaches (see, e.g., Ref.~\cite{Herman2023}) can be used.} combinatorial optimization problems~\cite{farhi2014quantum}. Given a binary\footnote{For any non-binary domains, a suitable encoding onto $\lbrace 0,1\rbrace^n$ must be conducted (see, e.g., Ref.~\cite{Tamura2021}).} objective function $f:\lbrace 0,1\rbrace^n\rightarrow\mathbb{R}$, this is accomplished by:
\begin{enumerate}
    \item Mapping the objective values onto the energy levels of a Hamiltonian $H_C=\sum_x f(x)\ket{x}\!\bra{x}$.
    \item Preparing a system in the ground state of a Hamiltonian, i.e., usually $\ket{+}^{\otimes n}$ for $H_M=-\sum_{i=1}^n \sigma_i^x$.
    \item Simulating the time evolution $\exp(i\int_0^T H_s(t)dt)$ approximatively, where $H_s(t)=\left(1 - s(t)\right)H_M + s(t)H_C$ governs the adiabatic evolution and $s:\left[0,T\right]\rightarrow\left[0,1\right]$ monotonically transitions from $0$ to $1$ for any given time $T>0$.
    \item Measuring the resulting state $\ket{\psi}$ and remapping it to its corresponding solution of the objective function $f$.
\end{enumerate}
To simulate the time evolution governed by $H_s$ on gate-based quantum computers, a discretization into $p\in\mathbb{N}$ Hamiltonians $H_s(1/T),...,H_s(T)$, as well as first order Suzuki-Trotter approximation is applied to yield the following unitary time evolution describing the QAOA:
\begin{align}
    U\left(\beta, \gamma\right) = U_M(\beta_p) U_C(\gamma_p) \ldots U_M(\beta_1) U_C(\gamma_1),
\end{align}
where $\beta_i$ and $\gamma_i$ characterize the speed of the time evolution and $U_M(\beta_i) = e^{-i\beta_i H_{M}}$, $U_C(\gamma_i) = e^{-i\gamma_i H_{C}}$, such that $U\left(\beta, \gamma\right)$ approaches adiabatic evolution for $p\rightarrow \infty$, and constant speed, i.e., $\beta_i = 1-i/p$, and $\gamma_i = i/p$~\cite{sack2021}. This is formally stated in \Cref{thm:QAOA}.

\begin{theorem}[Quantum Approximate Optimization Algorithm~\protect{\cite{farhi2014quantum,sack2021}}]\label{thm:QAOA}
    Given an objective function $f:\lbrace 0,1\rbrace^n\rightarrow\mathbb{R}$, the quantum circuit defined by $U\left(\beta, \gamma\right)$ in \Cref{eq:QAOA} yields $\argmin_x f(x)$ for $p\rightarrow \infty$, and $\beta_i \coloneqq 1-i/p$, and $\gamma_i \coloneqq i/p$.
    \begin{equation}\label{eq:QAOA}
        U\left(\beta, \gamma\right) \coloneqq U_M(\beta_p) U_C(\gamma_p) \ldots U_M(\beta_1) U_C(\gamma_1)H^{\otimes n},
    \end{equation}
    where $U_M(\beta_i) \coloneqq e^{-i\beta_i H_{M}}$, $U_C(\gamma_i) \coloneqq e^{-i\gamma_i H_{C}}$, $H_C\coloneqq\sum_x f(x)\ket{x}\!\bra{x}$, and $H_M\coloneqq -\sum_{i=1}^n \sigma_i^x$.
\end{theorem}

The key difference between pure adiabatic time evolution (as in the QAA) and the QAOA is the introduction of the parameters $\beta$ and $\gamma$, which allow for non-linear evolution speed. This is a crucial benefit, as the maximal evolution speed allowed by the adiabatic theorem at any given point in time $t$ depends on the spectral gap (i.e., the energy gap between the ground state and the first exited state of the continuous-time Hamiltonian $H(t)$), which is computationally intractable in the general case~\cite{farhi2014quantum}. Through the parameterization of this time evolution speed, machine learning techniques can be used to optimize it~\cite{Mitarai2018}, which can significantly increase the solution quality under restricted runtime~\cite{farhi2014quantum}.

Note that as proposed in Ref.~\cite{Bärtschi2020GroverMixers}, the QAOA can be adapted to handle constraints in form of a reduced space of feasible solutions $F\subseteq \lbrace 0,1\rbrace^n$ given by a state preparation operator $U_s:\ket{0}^{\otimes n}\mapsto \ket{F}$, where $\ket{F}$ a superposition of all feasible solutions.

\begin{lemma}[Grover Mixer~\protect{\cite{Bärtschi2020GroverMixers}}]
    By adapting $U_M(\beta_i) \coloneqq e^{-i\beta_i\ket{F}\bra{F}}$ and $U\left(\beta, \gamma\right)$ as in \Cref{eq:GMQAOA},
    \begin{equation}\label{eq:GMQAOA}
        U\left(\beta, \gamma\right) \coloneqq U_M(\beta_p) U_C(\gamma_p) \ldots U_M(\beta_1) U_C(\gamma_1)U_s,
    \end{equation}
    the space of explored solutions in the QAOA can be limited to $F\subseteq \lbrace 0,1\rbrace^n$. When $\ket{F}$ is an equal superposition over all states in $F$, solutions with the same solution quality are sampled with equal probability.
\end{lemma}

\subsection{Data input}
Efficient data input to quantum algorithms has been an intensive area of research in recent years. The key insight that emerged during this process is an efficient implementation of a uniformly controlled gate using arbitrary amounts of ancillary qubits~\cite{10044235}. This result allows for solving the well-known input problem in quantum computing by trading off runtime with space, i.e., the quantum statevector $\ket{\psi}$ of an arbitrary normalized vector $\psi\in\mathbb{R}^N$ can be prepared in time $\mathcal{O}(\log_2 N)$ if $\mathcal{O}(N)$ ancilla qubits are available, where $n\in\mathbb{N}$ s.t. $N=2^n$. As we will make use of this idea at many points in the paper, we now introduce these concepts more formally.

\begin{definition}[Uniformly Controlled Unitary~\protect{\cite{Yuan2023optimalcontrolled}}]\label{def:UCU}
    Given a collection of $2^k$ many $l$-qubit unitary matrices $U_0, U_1, ..., U_{2^k-1}$, we call the block-diagonal matrix $\textnormal{diag}(U_0,...,U_{2^k-1})$ a $(k,l)$-uniformly controlled unitary (UCU). If $l=1$, we use the term uniformly controlled gate (UCG) instead. A quantum circuit implementation of a $(k,l)$-UCU is displayed in \Cref{fig:UCU}.
\end{definition}

\begin{figure}[hbtp]
\centering
\input{tikz/UCU}
\caption{Quantum circuit implementation of a $(k,l)$-UCU as defined in \Cref{def:UCU}. The\begin{quantikz}\uctrl{}\end{quantikz}symbol is used as shorthand to iterate over all possible control combinations on the applied wire.}
\label{fig:UCU}
\end{figure}

\begin{lemma}[Implementing UCGs~\protect{\cite[Lem.~12]{10044235}}]\label{lem:UCG}
    Any $n$-qubit UCG can be implemented by a quantum circuit of depth $\mathcal{O}(n+\nicefrac{2^n}{n+m})$ using $m \in\mathbb{N}_0$ ancillary qubits.
\end{lemma}

\begin{lemma}[Quantum State Preparation~\protect{\cite[Thm.~1]{10044235}}]\label{lem:QSP}
    Any $n$-qubit quantum state $\ket{\psi}$ can be prepared by a quantum circuit of depth $\mathcal{O}(n)$ using $\mathcal{O}(2^n)$ ancillary qubits.
\end{lemma}

\begin{theorem}[Controlled Quantum State Preparation~\protect{\cite[Thm.~1]{Yuan2023optimalcontrolled}}]\label{thm:CQSP}
    For any set of $n$-qubit quantum states $\lbrace\ket{\psi_i}: i\in \lbrace 0,1\rbrace^k\rbrace$, a depth $\mathcal{O}\left(n+k+\nicefrac{2^{n+k}}{n+k+m}\right)$ quantum circuit performing the controlled state preparation $\ket{i}\ket{0}^{\otimes n}\mapsto \ket{i}\ket{\psi_i}$ $\forall i$ (i.e., a uniformly controlled unitary) can be implemented using $m \in\mathbb{N}_0$ ancillary qubits.
\end{theorem} 

\begin{remark}
    Let $2^k=K\in\mathbb{N}$ and $2^n=N\in\mathbb{N}$. When given access to $N\cdot K$ ancillary qubits, the algorithm provided by \Cref{thm:CQSP} enables the state preparation of $K$ normalized $N$-dimensional vectors $b_x\in\mathbb{R}^N$ $\forall x\in\lbrace 0,1\rbrace^k$ in a quantum circuit of depth $\mathcal{O}\left(\log_2(K) + \log_2(N)\right)$.
\end{remark}

\begin{lemma}[Polynomial Quantum Arithmetic~\protect{\cite[Appendix~B]{häner2018optimizingquantumcircuitsarithmetic}}]\label{lem:quantarithmetic}
    Given an $n$-qubit state $\ket{x}$ of a basis encoded binary number $x\in\left[-1,1\right[$, we can implement a unitary operator $\ket{x}\mapsto \ket{f(x)}$ in a quantum circuit of depth $\mathcal{O}(n^2d)$ using $\mathcal{O}(nd)$ ancillas for any bijective $d$-degree polynomial $f\in\mathbb{R}[x]$ with $|f(x)|\leq 1 \, \forall x$.
\end{lemma}

\begin{lemma}[Ancilla Quantum Encoding]\label{lem:AQE}
    Given a basis encoded quantum state $\ket{x}\coloneqq \ket{x_1...\, x_n}$ in two's complement (i.e., the binary string $x_1 ...\, x_n$ represents the number $x=-\frac{1}{2} x_1 + \sum_{i=2}^{n}x_i 2^{-i}\in \left[-0.5,0.5\right[$), we can implement ancilla quantum encoding (AQE), i.e., the unitary operator mapping $x$ onto an amplitude as in $\ket{0}\ket{x}\mapsto x \ket{0}\ket{x} + \sqrt{1-x^2}\ket{1}\ket{x}$, with a quantum circuit of depth $\mathcal{O}(n)$ and $\mathcal{O}(2^n)$ ancillary qubits or depth $\mathcal{O}(n^2/\varepsilon)$ and $\mathcal{O}(n/\varepsilon)$ ancillas, where $\varepsilon$ denotes the error.
\end{lemma}
\begin{proof}
    For the error dependent case see~\cite{10313905} and~\cite [Appendix B]{häner2018optimizingquantumcircuitsarithmetic}. The error-free case is based on a lookup table approach (cf.~\cite{10313905}) where each possible $x$ is processed using a respectively controlled $R_y$ rotation with classically precomputed rotation angles. This approach takes the form of a uniformly controlled gate, which can be implemented in the stated depth using $\mathcal{O}(2^n)$ ancillas based on \Cref{lem:UCG}.
\end{proof}

When solving an $A\vec{x}=\vec{b}$ type SLE on a quantum computer, one identifies the (necessarily normalized and potentially $0$-padded) vector $\vec{b}$ with a statevector $\ket{b}$, and the (necessarily $||\cdot||_2$-normalized) matrix $A$ with a unitary operator $U_A$ that block-encodes $A$, i.e.,
\begin{align}
    U_A=\left( {\begin{array}{cc}
   A & * \\
   * & *
  \end{array} } \right).
\end{align}
For any SLE that does not already take this form, the result of any quantum linear system solver for $x$ takes the (potentially $0$-padded) form $\ket{x}=A^+\vec{b}/||A^+\vec{b}\,||_2$. To simplify notation (and to avoid clutter), we assume sufficient normalization (and padding) for $A$, $\vec{b}$ and $\vec{x}$ respectively in the following if not indicated otherwise.

In the following we formally define the notion of block-encoding, and provide a concrete quantum circuit implementation to synthesise the corresponding unitary operator. Note that block-encoding implementations frequently entail a so called subnormalization factor, i.e., they block-encode $A/\alpha$  with $\alpha>1$ instead of $A$. As the circuit depth of state-of-the-art quantum linear system solvers typically depends linearly on $\alpha$ we focus on an implementation of block-encoding that minimizes subnormalization (cf. \Cref{lem:OBE}).
\begin{definition}[Block-encoding]
    For any matrix $A\in \mathbb{C}^{2^l\times 2^r}$, an ($n$+$a$)-qubit unitary $U_A$ is an $(\alpha,a,\varepsilon)$-block-encoding of $A$, if
    \begin{align}
        \left|\left| A-\alpha \left(\bra{0}^{\otimes a}\otimes I^{\otimes l}\right) U \left(\ket{0}^{\otimes a}\otimes I^{\otimes r}\right) \right|\right|\leq \varepsilon.
    \end{align}
\end{definition}

\begin{lemma}[Block-encoding sparse-access matrices~\protect{\cite[Lem.~48]{gilyen2018quantum}}]\label{lem:OBE}
    Assume sparse-access to an $s_r$-row- and $s_c$-column-sparse matrix $A\in \mathbb{R}^{2^n \times 2^n}$ with $|a_{ij}|\leq 1$, i.e., we have access to the oracles
    \begin{align*}
        O_r:\ket{i}\ket{k}&\mapsto\ket{i}\ket{r_{ik}} && \forall i\in \left[2^n-1\right], \, k\in\left[s_r\right]\\
        O_c:\ket{l}\ket{j}&\mapsto\ket{c_{lj}}\ket{j} && \forall l\in\left[s_c\right],\, j\in \left[2^n-1\right]\\
        O_A:\ket{i}\ket{j}\ket{0}^{\otimes e}&\mapsto\ket{i}\ket{j}\ket{\tilde{a}_{ij}} && \forall i,j\in \left[2^n-1\right]
    \end{align*}
    where $r_{ik}$ is the index of the $k$-th non-zero entry of the $i$-th row of $A$ and $k+2^n$ if there are less than $i$ non-zero entries, $c_{lj}$ is defined analogously, and $\tilde{a}_{ij}$ is a $\ceil{\log_{2}{1/\varepsilon_1}}$-bit binary approximation of $a_{ij}$ s.t. $\left|a_{ij}- \tilde{a}_{ij}\right|\leq\varepsilon_1$. Then we can implement a $(\sqrt{s_r s_c}, n+3, \varepsilon_1 + \varepsilon_2)$-block-encoding of $A$ using the quantum circuit described in \Cref{fig:blockencoding}, which requires a single use of $O_R$ and $O_C$, two uses of $O_A$, $\mathcal{O}(n+\log^{2.5}(s_r s_c/\varepsilon_2))$ one and two qubit gates and $\mathcal{O}(\log{1/\varepsilon_1},\log^{2.5}(s_r s_c/\varepsilon_2))$ ancillas. Here, $\varepsilon_2$ denotes the error resulting from an AQE of the values of the matrix entries.
\end{lemma}
\begin{proof}
    The only difference to the original formulation of this lemma in Ref.~\cite{gilyen2018quantum} is the added error-tolerance for the entries $\varepsilon_1$, which contributes practically linearly to the total error as the only function applied to the matrix entries (i.e., $\arccos$ during AQE) is basically linear near $0$. Thus the stated error dependence is slightly approximative, but accurate enough for our means, especially when assuming $\varepsilon_1$ to be small.
\end{proof}

\begin{figure}[hbtp]
\centering
\input{tikz/blockencoding}
\caption{Circuit of a $(\sqrt{s_r s_c}, n+3, \varepsilon_1+\varepsilon_2)$-block-encoding of a matrix $A$ given corresponding sparse-access oracles $O_r$, $O_c$, and $O_A$ as defined in \Cref{lem:OBE}. $D_s$ is defined as the map $\ket{0}^{\otimes q}\mapsto \frac{1}{\sqrt{s}}\sum_{k=1}^s \ket{k}$ (with $2\leq s \leq 2^q$ -- for an implementation requiring no ancillary qubits and a depth of $\mathcal{O}(\log_2 s)$ see Ref.~\cite{Shukla2024}. Purely for notational simplicity, AQE is visualized by a uniformly controlled $R_y$ rotation (cf. \Cref{lem:AQE}).}
\label{fig:blockencoding}
\end{figure} 

\begin{remark}
    In the case of $||A||_2\leq 1/2$, the subnormalization factor can be amplified to $\sqrt{2n_r n_c}$ (where $n_r\in\left[1,s_r\right]$ is an upper bound on $||a_{i\cdot}||_q^q$ and $n_c\in\left[1,s_c\right]$ is an upper bound on $||a_{\cdot j}||_{2-q}^{2-q}$ with $q\in\left[0,2\right]$) using uniform spectral gap amplification (for details, see \cite[Lem.~49]{gilyen2018quantum}), which we omit for ease of readability.
\end{remark}
\begin{corollary}
    As a consequence of \Cref{lem:OBE}, the speedup that we can gain (assuming a sufficiently well-conditioned SLE) from a quantum algorithm using this kind of block encoding depends on the sparsity. While an exponential speedup is possible for matrices whose sparsity is maximally logarithmic wrt. the systems dimensions, the worst case runtime for dense matrices scales with $\mathcal{O}(N)$.
\end{corollary}
\begin{proof}
    Acknowledge $s_r,s_c\leq N$ in \Cref{lem:OBE}, thus the subnormalization factor is upper bound by $N$.
\end{proof}

Note that this leads to a worst case quantum speedup of $\mathcal{O}(N^2)$ compared to the best classical approach (the Conjugate-Gradient method, which has complexity $\mathcal{O}(Ns\kappa\log(1/\varepsilon))$) when assuming a sufficiently well-conditioned SLE. If the condition number scales badly, we have to compare against the classical Gaussian elimination approach, which takes $\mathcal{O}(N^3)$ time.

\begin{remark}
    Given QRAM access to the matrix $A$ (for a definition of QRAM, see~\cite{qram})\footnote{While no hardware implementing QRAM is available at the time this article is published, one could use CQSP (cf. \Cref{thm:CQSP}) as a quantum circuit implementation of it at the cost of quadratically many ancillas wrt. the SLE's dimension.}, the circuit depth for a quantum linear systems solver for dense SLEs can be reduced to $\mathcal{O}(\kappa^2\sqrt{N} \polylog(N)/\varepsilon)$~\cite{Wossnig2018}. However, this approach is based on an entirely different algorithm (the Quantum Singular Value Estimation~\cite{Kerenidis2017}), which works based on quantum phase estimation and hence has an exponentially worse dependence on the error compared to QSVT. Note that this runtime can be improved to $\mathcal{O}(\kappa^2\sqrt{N} \polylog(\kappa/\varepsilon))$ if we also have access to an Linear Combination of Unitaries (LCU) decomposition of our matrix (which is generally not the case for practically relevant problems), by using the approach described in~\cite{wang2020quantum}. 
\end{remark}


\subsection{Quantum Linear Algebra Subroutines}
In this section, we show how Quantum Singular Value Transformation can be used to solve SLEs. Historically, QSVT is a generalization of Quantum Signal Processing (QSP). QSP is motivated from Ref.~\cite{PhysRevX.6.041067} and was later formally proposed in Ref.~\cite{PhysRevLett.118.010501}.
\begin{theorem}[Quantum Signal Processing~\protect{\cite[Thm.~5]{gilyen2018quantum}}]\label{thm:QSP}
    Given an operator $W(x)\coloneqq R_x(-2\arccos x)$ and a polynomial $P\in \mathbb{C}\left[x\right]$, we can find phase angles $\phi_1, ..., \phi_{d+1} \in \mathbb{R}$ and a polynomial $Q\in \mathbb{C}\left[x\right]$ such that
    \begin{equation*}
        e^{i\phi_1 \sigma_z}\prod_{k=1}^{d+1} W(x) e^{i\phi_k \sigma_z}=
        \begin{bmatrix}
          P(x) & i Q(x)\sqrt{1-x^2} \\
          i Q^*(x)\sqrt{1-x^2} & P^*(x) 
        \end{bmatrix}
    \end{equation*}  
    when $\tilde{P}\coloneqq \textnormal{Re}(P)$ satisfies 
    \begin{enumerate*}[label=(\roman*)]
        \item \label{itm:polycond1} $\textnormal{deg}(\tilde{P})\leq d$,
        \item \label{itm:polycond2} $\tilde{P}$ \textnormal{has parity} $d \mod 2$, and
        \item \label{itm:polycond3} $\tilde{P}(x)^2\leq 1$ 
    \end{enumerate*}, for all $x\in\left[-1,1\right]$.
\end{theorem}

\begin{proof}
    Special case ($\textnormal{Re}(Q)\equiv 0$) of Thm.~5 in Ref.~\cite{gilyen2018quantum}.
\end{proof}

\begin{remark}
    Usually, one is only interested in the real part of $P$ in \Cref{thm:QSP}. To acquire a unitary operation that performs $\textnormal{Re}(P(x))$ rather than $P(x)$, we can use that $\textnormal{Re}(P(x)) = \frac{1}{2}\left(P(x)+P^*(x)\right)$ in combination with the approach elaborated in Lem.~52 of Ref.~\cite{gilyen2018quantum} to implement such linear combination of unitary matrices as a $(1,1+1,0)$-block-encoding
    \begin{equation*}
        \begin{bmatrix}
          \textnormal{Re}(P(x)) & * \\
          * & * 
        \end{bmatrix} \ket{0}\ket{\psi}= \begin{quantikz}[column sep=5.3pt, row sep={25pt,between origins}]
\lstick{$\ket{0}$} & \gate{H} & \octrl{1} & \ctrl{1} & \gate{H} & \rstick{} \\
\lstick{$\ket{\psi}$}  & \qw & \gate[disable auto height]{U_\phi} & \gate[disable auto height]{U^\dagger_\phi} & \qw &  \rstick{} 
\end{quantikz}
    \end{equation*}
    where $U_\phi$ denotes the unitary operator proposed in \Cref{thm:QSP}, i.e., $U_\phi\coloneqq e^{i\phi_0 \sigma_z}\prod_{k=1}^d W(x) e^{i\phi_k \sigma_z}$. Note that this circuit also works for block encodings of arbitrary complex matrices $A$ instead of the scalar entry considered for $U_\phi$.
\end{remark}

\begin{remark}
    The state-of-the-art phase angle calculation approach for QSP (i.e., Ref.~\cite{dong2023robust}) is based on Newton's method and takes $\mathcal{O}(d^2)$ steps to converge in practice, which can pose severe limitations to highly ill-conditioned systems of equations. However, recently, a generalized from of QSP (GQSP) was proposed by Montlagh and Wiebe in Ref.~\cite{motlagh2024generalized} which allows for an $\mathcal{O}(d\log d)$ time algorithm to calculate the phase angles. The key to achieve this speedup is the generalization to arbitrary rotations in the signal processing operator by lifting the limitation of purely applying rotations in the $z$-axis. Based on this idea, Sünderhauf has proposed a corresponding generalization of the quantum singular value transformation algorithm (GQSVT), which uses the same quasi-linear approach to calculate the phase angles~\cite{sünderhauf2023generalized}. As the GQSVT has the same computational complexity as the QSVT apart from the faster phase angle calculation, we only discuss the more well-known QSVT in the following for the sake of brevity.
\end{remark} 

\begin{theorem}[Quantum Singular Value Transformation~\protect{\cite[Thm.~4]{PRXQuantum.2.040203}}]\label{thm:qsvt}
    Given a matrix  $A\in \mathbb{C}^{L\times R}$ with singular value decomposition $\sum_k \sigma_k \ket{w_k}\bra{v_k}$ through a corresponding $(\alpha,a,\varepsilon)$-block-encoding $U_A$, we can implement an $(\alpha,a,\varepsilon)$-block-encoding of $P(A)\coloneqq \sum_k P(\sigma_k) \ket{w_k}\bra{v_k}$ for any given odd polynomial $P\in \mathbb{C}\left[x\right]$ of degree $d$ satisfying conditions \ref{itm:polycond1} -- \ref{itm:polycond3} from \Cref{thm:QSP} via the unitary operator
    \begin{equation*}
        \tilde{\Pi}_{\phi_1} U_A \prod_{k=1}^{(d-1)/2}\left(\Pi_{\phi_{2k}} U_A^\dagger \tilde{\Pi}_{\phi_{2k+1}} U_A\right) \Pi_{\phi_{d+1}},
    \end{equation*}
    and respectively $P(A)\coloneqq \sum_k P(\sigma_k) \ket{v_k}\bra{v_k}$ for an even polynomial $P\in \mathbb{C}\left[x\right]$ of degree $d$ via
    \begin{equation*}
        \prod_{k=1}^{d/2}\left(\Pi_{\phi_{2k-1}} U_A^\dagger \tilde{\Pi}_{\phi_{2k}} U_A\right)  \Pi_{\phi_{d+1}},
    \end{equation*}
    where $\tilde{\Pi}_{\phi_k}\coloneqq e^{i\phi_k (2\tilde{\Pi} - \mathbb{I})}$ and $\Pi_{\phi_k}\coloneqq e^{i\phi_k (2\Pi - \mathbb{I})}$ denote the so-called projector-controlled phase shift operators with phase angles $\phi_k$ analog to \Cref{thm:QSP}. These are based on the orthogonal projectors (i.e., idempotent Hermitians) $\tilde{\Pi}\coloneqq \ket{0}^{\otimes m-l}\bra{0}^{\otimes m-l}\otimes I^{\otimes l}$ and $\Pi\coloneqq \ket{0}^{\otimes m-r}\bra{0}^{\otimes m-r} \otimes I^{\otimes r}$ for $m\coloneqq n+a$ that locate $A$ in its given block-encoding. Employing a single ancillary qubit for the implementation of the projector-controlled phase shift operators, we can implement the full QSVT procedure via the quantum circuits displayed in \Cref{fig:qsvt}.
\end{theorem}

\begin{figure*}[htb]
\centering
\begin{minipage}{\textwidth}
\centering
\input{tikz/qsvt-odd}\vspace{0.3cm}
(a)\label{fig:qsvt-odd} Quantum circuit corresponding to odd $P$.
\end{minipage}\\\hfill
\begin{minipage}{\textwidth}
\centering
\input{tikz/qsvt-even}\\\vspace{0.3cm}
(b)\label{fig:qsvt-even} Quantum circuit corresponding to even $P$.
\end{minipage}
\caption{Quantum circuits implementing an $(\alpha,a+1,\varepsilon)$-block-encoding of $P(A)$ via QSVT for arbitrary given $A\in \mathbb{C}^{L\times R}$ and even or odd polynomials $P\in \mathbb{C}\left[x\right]$. The blue boxes show implementations of projector-controlled phase shift operators $\tilde{\Pi}_{\phi_k}$ and $\Pi_{\phi_k}$ using a clean ancillary qubit from the top wire.}
\label{fig:qsvt}
\end{figure*}

\begin{remark}
     While the phase angles for QSP and QSVT are identical~\cite[Thm.~17]{gilyen2018quantum}, note that slightly altered versions of the proposed circuits exist in literature, which may demand a specific shift for each phase angle (for details, see Ref.~\cite{PhysRevA.103.042419}). For practical implementation, there is open source code available to calculate the phase angles. A state-of-the-art approach can be found in the Newton method from \href{https://github.com/qsppack/qsppack/}{QSPPACK} -- note though, that for the implementation to work with our definition of QSVT, one has to shift its first and last angle output by $-\pi/4$.
\end{remark}

\begin{theorem}[Quantum Moore-Penrose Pseudoinverse~\protect{\cite[Thm.~41]{gilyen2018quantum}}]\label{thm:matinv}
    Given $A\in \mathbb{C}^{L\times R}$ with singular value decomposition $A=W\Sigma V^\dagger$, an upper bound $\sigma_{\max}^{*}$ on its maximal singular value $\sigma_{\max}$ and a lower bound $\sigma_{\min}^{*}>0$ on its minimal non-zero singular value $\sigma_{\min}$, we can implement a $(2/\sigma_{\min}^{*},a+1,\varepsilon_2+\varepsilon_1/2\sigma^{*}_{\max} (\sigma^{*}_{\min} - \varepsilon_1))$-block-encoding of its pseudoinverse $A^{+}$ using QSVT on a $(\sigma_{\max}^{*},a,\varepsilon_1)$-block-encoding of $A$ with an odd polynomial of degree $\mathcal{O}(\kappa^{*} \log(\kappa^{*}/\varepsilon_2))$ that approximates ${\sigma_{\min}^{*}}/{2\sigma_{\max}^{*}x}$, where $\kappa^{*}\coloneqq \sigma^{*}_{\max}/\sigma^{*}_{\min}$, $A^+\coloneqq V\Sigma^{-1}W^\dagger$, and $\Sigma^{-1}$ denotes the element-wise inverse of the non-zero elements of $\Sigma$. For the error in the resulting block-encoding, it is assumed that $\varepsilon_1<< \sigma^{*}_{\min}$, which is an obvious constraint if one cares about all small singular values being inverted with high accuracy.
\end{theorem}
\begin{proof}
    This is a result of Thm.~41 from Ref.~\cite{gilyen2018quantum} when plugging in an $\varepsilon_1$-approximation of the (sub-)normalized version of $A$ as specified. The additive dependence of the error term $\varepsilon_1/2\sigma^{*}_{\max} (\sigma^{*}_{\min} - \varepsilon_1)$ in the resulting block-encoding of $A^+$ can be computed straightforwardly with the stated assumption of $\varepsilon_1<< \sigma^{*}_{\min}$ by focusing on the smallest singular value.
\end{proof}


\begin{corollary}[Quantum Linear System Solving~\protect{\cite{lin2022lecturenotesquantumalgorithms}}]\label{cor:QLS}
    Given a system of linear equations $A\vec{x}=\vec{b}$ with $A\in\mathbb{R}^{2^l \times 2^r}$, an upper bound $\sigma_{\max}^{*}$ on its maximal singular value, a lower bound $\sigma_{\min}^{*}>0$ on its minimal non-zero singular value as well as $U_A$, a $(\sigma_{\max}^{*},a,\varepsilon_1)$-block-encoding of $A$ and an oracle preparing $U_b\ket{0}^{\otimes l} = \ket{b}\coloneqq\vec{b}/||\vec{b}\,||_2$, we can implement a quantum circuit yielding an $(\varepsilon_2+\varepsilon_1/2\sigma^{*}_{\max} (\sigma^{*}_{\min} - \varepsilon_1))$-approximation of the quantum statevector $\ket{x}\coloneqq A^+ \vec{b}/||A^+\vec{b}\,||_2$ with a single initial query of $U_b$ and $\mathcal{O}(\kappa^{*} \log(\kappa^{*}/\varepsilon_2))$ subsequent sequential queries to $U_A$ using one ancillary qubit as described in \Cref{thm:matinv}. Note that for an extraction of $\ket{x}$, a $\ket{0}$ post-selection on the $a+1$ ancillary qubits is necessary. This post-selection has success probability $({\sigma_{\min}^{*}}\left|\left|A^+ \ket{b}\right|\right|_2/2)^2\in\left[(1/2\kappa^{*})^2,(1/2)^2\right]$ and can be amplified to a value $\geq 1/2$ using $\mathcal{O}(\kappa^*)$ rounds of amplitude amplification.
\end{corollary}
\begin{proof}
    The stated form of a quantum linear systems solver is a direct result of applying the Moore-Penrose inverse from \Cref{thm:matinv} onto $\ket{b}$. For details on the post-selection, see Ref.~\cite{lin2022lecturenotesquantumalgorithms}.
\end{proof}

\begin{remark}[Rescaling $\ket{x}$]\label{rem:rescalingx}
    As described in \Cref{cor:QLS}, our quantum linear system solver yields the quantum state $\ket{x}=A^+ \vec{b}/||A^+\vec{b}\,||_2$, which is merely a rescaled version of the actual result $\vec{x}=A^+ \vec{b}$. As a classical calculation of the factor $||A^+\vec{b}\,||_2$ would generally require solving the SLE, any quantum speedup would be lost by the classical overhead to compute this scaling factor. Fortunately, we can use Quantum Amplitude Estimation (QAE) on the state of the ancillas being in the $\ket{0}$ state to compute the actual value of $\sigma_{\min}^{*}\left|\left|A^+ \ket{b}\right|\right|_2/2$, i.e., the amplitude of the state $\ket{0}^{\otimes a}\ket{x}$ (a formal definition of QAE is given later in \Cref{thm:QAE}). The basis-encoded two's complement representation of this amplitude can then be used to properly rescale computations involving $\ket{x}$. Note that for calculating $||A^+\vec{b}\,||_2$ from $\left|\left|A^+ \ket{b}\right|\right|_2$ we can use the equality $||A^+\vec{b}\,||_2 = \left|\left|A^+ \ket{b}\right|\right|_2 ||\vec{b}\,||_2$ and the fact that $||\vec{b}\,||_2$ is necessarily known.
\end{remark}

\begin{remark}[Solving The Input Problem]\label{rem:inputproblem}
    Following the requirements of \Cref{cor:QLS}, the quantum speedup for solving SLEs with the stated algorithm relies heavily on the computational complexity of the block-encoding of $A$, as well as the state preparation of $\ket{b}$. While $\ket{b}$ can straightforwardly be prepared in time logarithmic to the dimensions of the SLE via \Cref{lem:QSP} at the cost of linearly many ancillaries, the block-encoding of $A$ is more intricate. The tool we will use to accomplish this for arbitrary sparse matrices is the block-encoding technique for sparse-access matrices described in \Cref{lem:OBE}. When the matrix is particularly structured, significantly more efficient alternative block-encodings might be possible (cf.~\cite{doi:10.1137/22M1484298}). The main insight facilitating our approach is that the necessary oracles can be implemented using the controlled quantum state preparation routine of \Cref{thm:CQSP}. For any $A$ of dimensions $\leq N=2^n$, this idea yields $(\sqrt{s_r s_c}\sigma_{\max}^{*}, n+3, \varepsilon_1+\varepsilon_2)$-block-encoding of $A$ of depth $\mathcal{O}(\log(N)+\log(1/\varepsilon_1)+\log^{2.5}(s_rs_c/\varepsilon_{2}))$ while requiring $\mathcal{O}(N+1/\varepsilon_1+\log^{2.5}(s_rs_c/\varepsilon_2))$ ancillary qubits. Here, $\ceil{\log_{2}{1/\varepsilon_1}}\in\mathbb{N}$ represents the number of bits needed for an $\varepsilon_1$-approximation of the binary representation of the matrix entries of the block-encoding $a_{ij}/\sigma_{\max}^{*}$, so that $1/\varepsilon_1 = \mathcal{O}(\min_{ij}(|a_{ij}|)/\sigma_{\max}^{*})$ if we assume the desired accuracy for even the smallest entry (e.g. $10^{-7}$ for single precision or $10^{-16}$ for double precision) as a fixed value corresponding to the multiplicative constant hidden in the $\mathcal{O}$-notation. This approach results in a quantum linear system solver of depth $\mathcal{\tilde{O}}(\log(N)\sqrt{s_rs_c}\kappa^{*}\log{(\kappa^{*}\sqrt{s_rs_c}/\varepsilon_3)}\log(1/\varepsilon_1))$, $\mathcal{O}(n)$ working qubits and $\mathcal{\tilde{O}}(N + 1/\varepsilon_1)$ ancillaries to produce an $\mathcal{O}(\varepsilon_1+\varepsilon_2+\varepsilon_3)$-approximation of $\ket{x}$ that can be post-selected with success probability $\Omega((\kappa^{*}\sqrt{s_rs_c})^2)$, where $\varepsilon_3$ denotes the accuracy of the inversion polynomial.
\end{remark}

\subsection{Amplitude Arithmetic}
In this section, we will address non-linear amplitude transformations as in $\ket{\psi}=\sum_{j=0}^{2^n-1}\alpha_j \ket{j}\mapsto 1/\mathcal{N} \sum_{j=0}^{2^n-1}f(\alpha_j) \ket{j}$, for polynomially approximable functions $f\in\mathbb{R}\left[x\right]$, where $\mathcal{N}$ denotes the corresponding normalization factor. This allows for non-linear post processing steps for the output of the simulation problem.

\begin{theorem}[Diagonal block-encoding of state vectors~\protect{\cite[Thm.~2]{rattew2023nonlinear}}]\label{thm:diagonal-blockencoding}
    Given access to an $n$-qubit unitary $U:\ket{0}^{\otimes n}\mapsto\ket{\psi}$, the quantum circuit in \Cref{fig:diagonal-blockencoding} implements a $(1,n+2,0)$-block-encoding of the diagonal matrix $\textnormal{diag}(\textnormal{Re}(\ket{\psi}))$ in circuit depth $\mathcal{O}(n)$ and $\mathcal{O}(1)$ queries of a controlled-$U$ gate for $p=0$. The circuit can alternatively block-encode the imaginary part of $\ket{\psi}$ when setting $p=1$.
\end{theorem}

\begin{figure}[hbtp]
\centering
\input{tikz/diagonal-blockencoding}
\caption{Quantum circuit implementing a $(1,n+2,0)$-block-encoding of the diagonal matrix $\textnormal{diag}(\textnormal{Re}(\ket{\psi}))$ as defined in \Cref{thm:diagonal-blockencoding}. Implementations for the operators $G_p$ and $W_p$ are displayed in \Cref{fig:diagonal-blockencoding-Gp} an \Cref{fig:diagonal-blockencoding-Wp} in the Appendix.}
\label{fig:diagonal-blockencoding}
\end{figure}

\begin{theorem}[Non-linear amplitude arithmetic~\protect{\cite[Thm.~4]{rattew2023nonlinear}}]\label{thm:amplitude-arithmetic}
    Given access to an $n$-qubit unitary $U:\ket{0}^{\otimes n}\mapsto\ket{\psi}\in\mathbb{R}^{2^n}$, we can prepare an $\varepsilon$-approximation of $f(\ket{\psi})/\mathcal{N}$ for any given $f\in\mathbb{R}\left[x\right]$ satisfying $f(0)=0$, for which a degree $k=K(\varepsilon \mathcal{N}^2/\gamma 2^n)$ polynomial $P\in\mathbb{R}\left[x\right]$ exists, that $\frac{\gamma 2^n}{\varepsilon \mathcal{N}^2}$-approximates $f$ with $P(0)=0$, where $\gamma\coloneqq \max_{-1 \leq x \leq 1} \left|f(x)\right|$, $\mathcal{N}^2\coloneqq \left|\left|f(\ket{\psi})\right|\right|_2$, and $K$ denotes a function describing the degree of the polynomial in terms of the error. This can be done at arbitrarily high success probability, $\mathcal{O}(\tilde{\gamma} k/\mathcal{N})$ query complexity, an overall circuit depth of $\mathcal{O}(n\tilde{\gamma} k/\mathcal{N})$ and using $\mathcal{O}(n)$ ancillas, where $\tilde{\gamma}\coloneqq \max_{-1 \leq x \leq 1} \left|P(x)/x\right|$.
\end{theorem}

\subsection{Data readout}\label{subsec:data-readout}
In this section we present different subroutines that allow us to retrieve summary statistic information from an $n$-qubit quantum state $\ket{\psi}$ given its state preparation unitary $U:\ket{0}^{\otimes n}\mapsto \ket{\psi}$.
\begin{theorem}[Quantum Phase Estimation~\protect{\cite{kitaev1995quantum} and \cite[Appendix C]{cleve1998QAE}}]
    Given an $n$-qubit unitary operator $U$ and a corresponding eigenvector $\ket{\psi}$ s.t. $U\ket{\psi}=e^{2\pi i\theta}\ket{\psi}$ and wlog $\theta\in\left[-0.5,0.5\right[$, we can compute a basis-encoded $\varepsilon$-approximation of $\theta$ with success probability $1-\delta$ for any $0<\delta<1$ in a quantum circuit of query complexity $\mathcal{O}(1/\varepsilon\delta)$ using $n + \left(m+\ceil{\log_2\left(1/2\delta + 1/2\right)}\right)$ qubits with $m\coloneqq\ceil{\log_2\left(1/\varepsilon\right)}$. The additive time complexity on top of the stated query complexity is $1+ (\ceil{\log_2\left(1/\varepsilon\right)} + \ceil{\log_2\left(1/2\delta + 1/2\right)})^2$.
\end{theorem}

\begin{remark}
    The key for achieving the success probability of $1-\delta$, is doing a standard QPE involving all ancillary qubits, but then only using the result of the $\ceil{\log_2\left(1/\varepsilon\right)}$ most significant bits.
\end{remark}

\begin{theorem}[Quantum Amplitude Estimation~\protect{\cite[Thm.~12]{brassard2002quantum}}]\label{thm:QAE}
    Given an $n$-qubit unitary operator $\mathcal{A}:\ket{0}^{\otimes n}\mapsto \ket{\psi}\coloneqq e^{i\varphi}\cos{\theta}\ket{\psi_0} + \sin{\theta}\ket{\psi_1}$ with $-\pi/2\leq \theta \leq \pi/2$, we can compute a basis-encoded $(\varepsilon_1+\varepsilon_2)$-approximation of $\left|\sin\theta\right|$  with success probability $1-\delta$ for any $0<\delta<1$ by applying QPE on $\ket{\psi}$ and $\mathcal{Q}\coloneqq -\mathcal{A}U_0 \mathcal{A}^\dagger U_{\psi_1}$, where $U_0\coloneqq \mathbb{I} - 2\ket{0}^{\otimes n}\bra{0}^{\otimes n}$ and $U_{\psi_1}\coloneqq \mathbb{I} - 2\ket{\psi_1}\bra{\psi_1}$, and arithmetically post-processing the result with the function $\sin(\pi|\cdot |)$. Here $\varepsilon_2$ denotes the error of a polynomial approximation of the $\sin(\pi|\cdot |)$ function. Note that as the computational overhead of this arithmetic operation is fairly small in practice, we assume it to be constant in the rest of this paper for the sake of readability.
\end{theorem}
\begin{proof}
    Applying QPE on the specified operator $\mathcal{Q}$ with the initial state $\psi$ prepared using the given operator $\mathcal{A}$, yields an $\varepsilon_1$-approximation of $i/\sqrt{2}\left(e^{i\theta}\ket{\theta_{-}}\ket{\psi_-} - e^{-i\theta}\ket{\theta_{+}}\ket{\psi_+}\right)$, where $\theta_{\pm}\coloneqq\pm\nicefrac{\theta}{\pi}$ is encoded in twos-complement representation and $\ket{\psi_{\pm}}\coloneqq 1/\sqrt{2}(\ket{\psi_1}\pm i \ket{\psi_0})$. Using the well-known procedure to flip the sign of a bit string encoded in twos-complement (i.e., flipping all bits and then adding one) conditionally for $\ket{\theta_-}$ via an extra ancilla, one can perform the absolute value function. For executing the $z\mapsto\sin(\pi z)$ function, we employ a truncated series approximation of degree $d$ via \Cref{lem:quantarithmetic} which introduces an error $\varepsilon_2=\mathcal{O}(1/d!)$ additively on top of $\varepsilon_1$.
\end{proof}

\begin{remark}
    Perhaps the most common application of QAE involves the estimation of a single amplitude of $\ket{\psi}$, i.e., $\psi_1\in\left\lbrace 0, 1\right\rbrace^n$. In this case, $U_{\psi_1}$ can be implemented as a multi-controlled $Z$-gate sandwiched with an $X$- or $I$-gate on every $i$-th qubit, depending on whether the $i$-th bit of $\psi_1$ is $1$ or $0$ respectively. Note that the negative sign in $\mathcal{Q}$ can, e.g., be implemented by the gate sequence $I^{\otimes n-1} \otimes XZXZ$, which becomes necessary, as QPE applies $\mathcal{Q}$ as a controlled gate, s.t. its global phase matters.
\end{remark}

\begin{lemma}[Expectation Value via Hadamard Test~\protect{\cite{liao2021quantum}}]\label{lem:exp-val}
    Given $\ket{\psi}$ via $U:\ket{0}^{\otimes n}\mapsto\ket{\psi}$, we can compute an $(\varepsilon_1+\varepsilon_2)$-approximation of $\braket{\psi|H|\psi}$ with success probability $1-\delta$ for any $0<\delta<1$ by applying QAE on $\mathcal{Q}\coloneqq \mathcal{A}U_0 \mathcal{A}^\dagger U_{\psi_1}$, where $U_0\coloneqq \mathbb{I} - 2\ket{0}^{\otimes n}\bra{0}^{\otimes n}$, $U_{\psi_1}\coloneqq Z\otimes\mathbb{I}$, and $\mathcal{A}$ representing a Hadamard test of $\ket{0}^{\otimes m}\ket{\psi}$ with $U_H$, a given $(1,m,0)$-block-encoding of the hermitian $2^n \times 2^n$ matrix $H$. A quantum circuit implementing $\mathcal{Q}$ is displayed in \Cref{fig:exp-val}.
\end{lemma}
\begin{proof}
    Analog to \Cref{thm:QAE}, QPE yields a basis encoded superposition of $\pm\nicefrac{1}{\pi}\arcsin(\sqrt{1/2+\braket{\psi|H|\psi}/2})$ in twos-complement representation. Following the same steps as in the proof of \Cref{thm:QAE} for computing the function $z\mapsto \sin(\pi|z|)$, we then have to subsequently compute $z\mapsto 2(z^2 - 1/2)$, again by first computing the inner part by squaring $z$, then subtracting $1/2$ and finally multiplying by $2$. The error dependence is analog to \Cref{thm:QAE}.
\end{proof}

\begin{figure}[hbtp]
\centering
\input{tikz/exp-val}
\caption{Quantum circuit implementing $\mathcal{Q}$ from \Cref{lem:exp-val}.}
\label{fig:exp-val}
\end{figure}

\begin{corollary}[Fidelity via Hadamard Test~\protect{\cite{liao2021quantum}}]\label{cor:fid}
    Given $U_\psi:\ket{0}^{\otimes n} \mapsto \ket{\psi}$ and $U_\varphi:\ket{0}^{\otimes n} \mapsto \ket{\varphi}$, we can compute an $(\varepsilon_1 + \varepsilon_2)$-approximation of $\left|\braket{\varphi|\psi}\right|$ with success probability $1-\delta$ for any $0<\delta<1$ by applying the algorithm proposed in \Cref{lem:exp-val} to $m=0$, $U=I^{\otimes n}$ and $U_H\coloneqq U^\dagger_\varphi U_\psi$.
\end{corollary}
\begin{proof}
    Analog to \Cref{lem:exp-val}, QAE for the specified inputs yields $\pm\nicefrac{1}{\pi}\arcsin(\sqrt{1/2+\left|\braket{\varphi|\psi}\right|/2})$. The rest of the proof is completely analog to the one of \Cref{lem:exp-val}.
\end{proof}

\begin{remark}
    If one only has access to an oracle preparing a state $\ket{\psi\rq}=\sin\theta \ket{0}^{\otimes m}\ket{\psi} + e^{i\varphi}\cos{\theta}\ket{*}\ket{*}$ instead of $\ket{0}^{\otimes m}\ket{\psi}$, \Cref{lem:exp-val} and \Cref{cor:fid} can be applied nevertheless by adding $-Z$-gates to each non-zero qubit of the $m$-qubit register in the $U_{\psi_1}$ operator.
\end{remark}


\section{Methodology}
\label{sec:methodology}
In this section, we present a framework of quantum algorithms to solve LinQuSO problems. The general setup consists of the QAOA being used to solve the optimization, the QSVT for solving the SLE and the QPE for extracting summary statistic information. We start by proposing a straightforward quantum algorithm that allows for constructing the cost unitary $\ket{x}\mapsto e^{-i \gamma_i f(x)}$ given access to the circuit computing the summary statistic result of the simulation problem $\textnormal{QSim}\ket{x}\ket{0}^{\otimes m}=\ket{x}\ket{u(s(x))}$ to integrate the simulation component into the QAOA.

\begin{lemma}[Quantum Phase Application]\label{lem:QPA}
    Given a number $a\in \mathbb{R}$ in an encoding of the form $\sum_{i=1}^{m} \alpha_i a_i$, where $\alpha_i\in\mathbb{R}$ and $a_i\in\left\lbrace0,1\right\rbrace$ for all $i\in\left[m\right]$, we can implement $U_c:\ket{a}\mapsto e^{-i \gamma a}\ket{a}$ for any $\gamma\in\mathbb{R}$ via the depth-$1$ layer of phase gates $\bigotimes_{i=1}^{m}P(-\gamma \alpha_i)$.
\end{lemma}
\begin{proof}
    The result directly follows from the definition of the phase gate, as $\left(\bigotimes_{i=1}^{m}P(-\gamma \alpha_i)\right)\ket{a} = \bigotimes_{i=1}^{m}P(-\gamma \alpha_i)\ket{a_i}=e^{-i\gamma \sum_i \alpha_i a_i}\ket{a}=e^{-i \gamma a}\ket{a}$. Note however, that $\gamma$ has to be chosen small enough wrt. $a$, if the conducted phase application should be injective.
\end{proof}

For simplicity, we initially assume that the QuSO problem is already binary, unconstrained, and that its costs are already fully determined by the simulation problem, i.e., $f(x)=u(s(x))$. Possible generalizations are shown subsequently.

\begin{theorem}[QuSO Solver Architecture]\label{thm:quso}
    Given a quantum circuit implementing $\textnormal{QSim}\ket{x}\ket{0}^{\otimes m}=\ket{x}\ket{u(s(x))}$, we can implement the cost unitary $U_C(\gamma_i)\ket{x}=e^{-i \gamma_i u(s(x))}\ket{x}$ using $m$ ancillary qubits, one application of $\textnormal{QSim}$, the quantum phase application (QPA) as proposed in \Cref{lem:QPA}, and one application of $\textnormal{QSim}^\dagger$. Given user-specifiable oracles for the initial state preparation $U_S$ (e.g., $H^{\otimes n}$) and a mixer unitary (e.g., $R_X(-\beta_i)^{\otimes n}$), we can construct a quantum circuit implementing the QAOA for finding $\argmin_x u(s(x))$. For computing the costs of the final output (which is assumed to be severely less efficient classically), we add an application of $\textnormal{QSim}$ directly before the measurement to yield these costs. The complete circuit is displayed in \Cref{fig:quso}. For the sake of simplicity, we excluded a rescaling component from this construction, which could however be straightforwardly added via the implementation stated in \Cref{rem:rescalingx}.
\end{theorem}
\begin{proof}
    The following calculation proves that our 
    implementation of the cost unitary is correct.
    \begin{align*}
        &\textnormal{QSim}^\dagger\left(I^{\otimes n}\otimes\textnormal{QPA}(\alpha, \gamma_i)\right)\textnormal{QSim}\ket{x}\ket{0}^{\otimes m}\\
        =&\textnormal{QSim}^\dagger\left(I^{\otimes n} \ket{x} \otimes\textnormal{QPA}(\alpha,\gamma_i)\ket{u(s(x))}\right)\\
        =&\textnormal{QSim}^\dagger e^{-i \gamma_i u(s(x))}\ket{x}\ket{u(s(x))}\\
        =&e^{-i \gamma_i u(s(x))}\ket{x}\ket{0}^{\otimes m}
    \end{align*}
    As the rest of our implementation besides the extra measurement of $\ket{u(s(x))}$ is analog to a standard QAOA procedure (cf. \Cref{thm:QAOA}), this completes the proof.
\end{proof}

\begin{figure*}[hbtp]
\centering
\input{tikz/quso}
\caption{Quantum circuit solving a QuSO problem of the form $\argmin_x f(x)=u(s(x))$ as proposed in \Cref{thm:quso}. A potential implementation for QSim can be found in \Cref{fig:quso-U_f}. The\begin{quantikz}\ground{}\end{quantikz}symbol is used to mark ancillas in the $\ket{0}$ state that can be reused for further calculations.}
\label{fig:quso}
\end{figure*}

We now discuss how more general QuSO problems than the minimal form of $\argmin_x u(s(x))$ can be solved by extending the circuit structure presented in \Cref{thm:quso}.

\begin{remark}[Solving Mixed-Integer QuSO problems]\label{rem:MILPQuSO}
    As typical for any quantum optimization algorithm based on identifying the optimization problems' cost landscape with the energy spectrum of a Hamiltonian, we also have to resort to an appropriate, problem-dependent discretization of all continuous decision variables.
\end{remark}

\begin{remark}[Integrating arbitrary continuous cost functions]\label{rem:contquso}
    As a corollary of \Cref{rem:MILPQuSO}, the domain space of the decision variables is necessarily bounded. By defining a closed interval that contains this reduced domain space, we can make use of the Stone–Weierstrass theorem~\cite{stone1937applications} to find a polynomial approximation of the cost function $f$. This polynomial can be translated into an Ising Hamiltonian (cf., e.g.,~\cite{10.1145/3583133.3596358}) so that the value of the specific function can be computed in basis encoding using \Cref{lem:exp-val}. The summary statistic result of the simulation problem $u(s(x))$ can then be integrated into the rest of $f(x)$ via the QSim operator corresponding to $u(s(x))$ and quantum arithmetic operations. In the special case where $u(s(x))$ is just an additive term in $f(x)$, one can simply append the cost unitary corresponding to $u(s(x))$ to any previous cost unitaries of other terms of $f(x)$ in the quantum crircuit, as all cost operators commute.
\end{remark}

\begin{remark}[Integrating constraints into the QuSO solver]
    One possibility of integrating constraints in the QuSO solver architecture of \Cref{thm:quso} is via the introduction of penalty terms. This way, we can exploit the same approach as in \Cref{rem:contquso}. Details about the general approach to formulate constraints as penalty terms can be found in \Cref{subsec:quantopt}
\end{remark}

Having shown concrete techniques for solving general QuSO problems, we now focus on the implementation of QSim for a large class of LinQuSO problems. In particular, we prove that this approach allows for exploiting exponential quantum speedup for the simulation component. An overview of the framework of quantum algorithms we provide for implementing QSim is given in \Cref{tab:qblas} in the Appendix.

\subsection{Data input}\label{subsec:meth:data-input}
In this section, we show how the controlled quantum state preparation algorithm from \Cref{thm:CQSP} can be adapted to implement the quantum data input necessary for solving a decision variable dependent SLE $A_x \vec{y} = \vec{b}_x$ with the QSVT. The notation with the binary decision variables $x\in\left\lbrace 0,1 \right\rbrace ^n$ as a subscript to $A$ and $\vec{b}$ denotes that their entries depend on $x$. We now show how to efficiently prepare $\vec{b}_x$ and block-encode $A_x$ depending on their dependence on $x$.

\begin{lemma}[Fast controlled State Preparation]\label{lem:FCSP}
    Given a fixed $x\in\left\lbrace 0,1\right\rbrace^m$ and a normalized vector $\vec{b}_x\in\mathbb{R}^{2^n}$ with entries of arbitrary form, i.e., ${(\vec{b}_x)}_i=b(x)_i$ for all $x\in\left\lbrace 0,1\right\rbrace^m$ and $i\in\left[N\right]$ for arbitrary functions $b:\left\lbrace 0,1\right\rbrace^m\rightarrow \mathbb{R}^{2^n}$, we can implement a state preparation of $\vec{b}_x$ with a circuit of depth $\mathcal{O}(n)$ using $\mathcal{O}(N2^m)$ ancillary qubits.
\end{lemma}
\begin{proof}
    By applying \Cref{thm:CQSP} to execute $\ket{x}\ket{0}^{\otimes n}\mapsto \ket{x}\ket{b_x}$ one directly gets the stated complexities.
\end{proof}

In practice, the function $b(x)$ from \Cref{lem:FCSP} is usually less arbitrary, which can allow for exponentially better space requirements as shown in \Cref{thm:QDAC-bitstrings}. 

\begin{theorem}[Quantum Digital to Analog Conversion for Bitstrings]\label{thm:QDAC-bitstrings}
    Given an $n$-qubit computational basis state $\ket{x}$ with $n=2^\eta$ for some $\eta \in\mathbb{N}$ and a state preparation oracle $U_c:\ket{0}^\eta\mapsto \ket{c}$, we can perform $\ket{x}\ket{0}\ket{0}^{\otimes \eta}\mapsto \ket{x} \left(\ket{0}\sum^{\eta}_i c_ix_{k(i)} \ket{i} + \ket{1}\sum^{\eta}_i c_i(1-x_{k(i)}) \ket{i}\right)$ using a quantum circuit of depth $\mathcal{O}(\log n)$ and $\mathcal{O}(n^2)$ ancillary qubits, where $k:\left[n\right]\rightarrow\left[n\right]$ maps every index of $\ket{c}$ to an index of $\ket{x}$.
\end{theorem}
\begin{proof}
    A naive setup for constructing a circuit that allows for implementing the specified operator with $k(i)=i$ is displayed in \Cref{fig:QDAC-bitstrings}. As evident by the circuit construction, the core operation of the circuit is the $(\eta, n+1)$-UCU with $U_i$ defined as a controlled not gate with the target being fixed on the ancillary qubit and the target being on the $k(i)$-th qubit of the $\ket{x}$ register. Through an equivalent reformulation of this circuit, we convert the stated UCU into a much less complex UCU that allows the application of Lem. 10 from Ref.~\cite{Yuan2023optimalcontrolled} to implement this UCU\footnote{Note that the necessary conditions for Lem. 11 from Ref.~\cite{Yuan2023optimalcontrolled} are not satisfied in the construction of \Cref{fig:QDAC-bitstrings}, as the $U_i$ operators do not fulfill the property of being a \emph{standard quantum circuit}.}. For the reformulation, we use these two tricks: (1) control and target of a controlled not gate can be switched by a Hadamard sandwich, and (2) a controlled gate can be constructed via a sandwich of controlled swap operators (cf. \cite[Fig.~5]{PRXQuantum.2.040203}). This yields the circuit displayed in \Cref{fig:QDAC-Wx}, which clearly allows the application of Lem. 10 from Ref.~\cite{Yuan2023optimalcontrolled} and hence yields the unitary operator specified in the theorem with the stated complexities up to the implementation of the controlled swap gates. As the controlled swap gates on both sides result in permutation matrices, we employ \Cref{lem:permuations} to complete the proof.
\end{proof}

\begin{corollary}\label{cor:stateprep_of_arbitrary_bx}
    \Cref{thm:QDAC-bitstrings} can be generalized to $\eta$ being independent of $n$ by adapting the employed $(n,\eta)$-UCU with $k:\left[2^\eta\right]\rightarrow\left[n\right]$ while keeping $U_i=X_{k(i)}$. Using the same proof structure as in \Cref{thm:QDAC-bitstrings}, this generalization yields a circuit depth of $\mathcal{O}(\log n + \log N)$ and requires $\mathcal{O}(nN)$ ancillary qubits for $b_x\in\mathbb{R}^{N}$ for an $n$-qubit computational basis state $\ket{x}$.
\end{corollary}

\begin{figure}[hbtp]
\centering
\input{tikz/QDAC-bitstrings}
\caption{Quantum circuit for QDAC with bitstrings as shown in \Cref{thm:QDAC-bitstrings} for $k(i)=i$. The uniformly controlled unitary can be constructed efficiently using \cite[Lem. 10]{Yuan2023optimalcontrolled} on the UCU of the standard quantum circuit shown in \Cref{fig:QDAC-Wx}.}
\label{fig:QDAC-bitstrings}
\end{figure}

\begin{figure*}[hbtp]
\centering
\input{tikz/QDAC-Wx}
\caption{Standard quantum circuit realizing \Cref{thm:QDAC-bitstrings} for $k(i)=i$. $X_{k(i)}$ is chosen to be the $X$-gate only for the control being the bitstring $i-1$, otherwise it is asserted to the identity gate. For $k(i)\neq i$, the swaps have to be carried out on the $k(i)$-th qubits of the top and bottom registers.}
\label{fig:QDAC-Wx}
\end{figure*}

\begin{lemma}[Fast Permutation Operators]\label{lem:permuations}
    Given a sequence of $n$ swap operations $\textnormal{S}_i\coloneqq\textnormal{SWAP}(j_i,k_i)$ on $n$ qubits with $j_i,k_i\in\left[n\right]\,\forall i\in\left[n\right]$, we can implement the resulting unitary operator $\prod_i \textnormal{S}_{i\in \left[n\right]}$ in depth $\mathcal{O}(\log n)$ using $\mathcal{O}(1)$ ancillary qubits.
\end{lemma}
\begin{proof}
    We can classically precompute the resulting permutation in $\mathcal{O}(n)$ steps and then use \Cref{lem:OBE} together with \Cref{thm:CQSP} to achieve the stated complexity. As the block-encoded matrix is unitary, no post-selection is necessary.
\end{proof}



We now continue by showing how to accomplish similar complexities for matrix block-encoding. For that, we start with the baseline of block-encoding a matrix independent of any decision variables by formalizing \Cref{rem:inputproblem}.

\begin{lemma}[Fast Block-Encoding]\label{lem:FOBE}
    Given an $s_r$-row- and $s_c$-column-sparse matrix $A\in\mathbb{R}^{2^n \times 2^n}$ as well as an upper bound $\sigma_{\max}^{*}$ on its maximal singular value, we can implement a $(\sqrt{s_rs_c}\sigma_{\max}^{*},n+3,\varepsilon_1+\varepsilon_2)$-block-encoding of $A$ with circuit depth $\mathcal{\tilde{O}}(n)$ using $\mathcal{\tilde{O}}(N+\log 1/\varepsilon_1)$ ancillary qubits, where $1/\varepsilon_1=\mathcal{O}(\sigma_{\max}^{*}/\min_{ij}|a_{ij}|)$.
\end{lemma}
\begin{proof}
    We use the general circuit architecture of \Cref{lem:OBE} and \Cref{thm:CQSP} for implementing each oracle to yield the desired unitary operator with the stated complexities.
\end{proof}

\begin{lemma}[Fast controlled Block-Encoding]\label{lem:FCOBE}
    Given $A_x\in\mathbb{R}^{2^n \times 2^n}$ with $s_r$ denoting the minimal row- and $s_c$ the minimal-column-sparsity of $A_x$ over all $x$ as well as an upper bound $\sigma_{\max}^{*}$ on the maximal singular value of all $A_x$, we can implement a $(\sqrt{s_rs_c}\sigma_{\max}^{*},n+3,\varepsilon_1+\varepsilon_2)$-block-encoding of $A_x$ for every $x\in\left\lbrace 0,1\right\rbrace^m$ with circuit depth $\mathcal{\tilde{O}}(n)$ using $\mathcal{O}(N2^m+\log 1/\varepsilon_1)$ ancillary qubits where $1/\varepsilon_1=\mathcal{O}(\sigma_{\max}^{*}/\min_{ij}|a_{ij}|)$.
\end{lemma}
\begin{proof}
    Analog to \Cref{lem:FCSP}, we use the general circuit architecture of \Cref{lem:OBE} and \Cref{thm:CQSP} for implementing each oracle, however, to account for the fact that the oracles are now dependent on $x$, we have to extend the control register in \Cref{thm:CQSP} to include $x$ which yields the increased complexity.
\end{proof}

\begin{theorem}[Optimal controlled block-encoding]\label{thm:OCOBE}
    Given $A_x\in\mathbb{R}^{2^n \times 2^n}$ with $s_r$ denoting the minimal row- and $s_c$ the minimal-column-sparsity of $A_x$ over all $x$ with entries of the form $a_{ij}x_{k(i,j)}$ where $k:\left[n\right]^2\rightarrow\left[m\right]$ maps every entry index of $A$ to an index of $x$, as well as an upper bound $\sigma_{\max}^{*}$ on the maximal singular value of all $A_x$, we can implement a $(\sqrt{s_rs_c}\sigma_{\max}^{*},n+3,\varepsilon_1+\varepsilon_2)$-block-encoding of $A_x$ for every $x$ with circuit depth $\mathcal{\tilde{O}}(n)$ using $\mathcal{O}(Nm+\log 1/\varepsilon_1)$ ancillary qubits where $1/\varepsilon_1=\mathcal{O}(\sigma_{\max}^{*}/\min_{ij}|a_{ij}|)$.
\end{theorem}
\begin{proof}
    The exponentially reduced space complexity in $m$ can be achieved by a similar setup as in \Cref{thm:QDAC-bitstrings}. More concretely, we start with our typically employed block-encoding from \Cref{lem:FOBE} for a matrix with entries $a_{ij}$ but stop before the AQE step. Before conducting the AQE and $O_A^\dagger$, we insert the UCU from \Cref{thm:QDAC-bitstrings} controlled on the matrix index registers to compute the value of the respective $x_{k(i,j)}$ on the ancillary qubit. Then we control the AQE on this ancilla to make sure that the value for $a_{ij}$ is only respected if the corresponding $x_{k(i,j)}$ equals one. After that, we uncompute this step (except for the AQE) to then finally uncompute $O_A$ to finish the block-encoding. By applying the implementation setup from \Cref{fig:QDAC-Wx} for the UCU, we end up with an algorithm of the stated complexity.
\end{proof}

\subsection{Amplitude Arithmetic}
In this section, we show an upper bound on the complexity of performing the absolute value function on quantum amplitudes as an addendum to list of non-linear amplitude manipulation functions proposed in Thm.~5 of Ref.~\cite{rattew2023nonlinear}. 
\begin{lemma}[Approximating $|\cdot|$]\label{lem:absval}
    Given access to an $n$-qubit unitary $U:\ket{0}^{\otimes n}\mapsto\ket{\psi}\in\mathbb{R}^{2^n}$, we can implement an $\varepsilon$-approximation of state preparation unitary for $\ket{|\psi|}$ via \Cref{thm:amplitude-arithmetic} using the polynomial
    \begin{equation}
        P(x)=\frac{2}{\pi} + \frac{4}{\pi}\sum_{k=1}^d \frac{(-1)^{k+1}}{4k^2-1}T_{2k}(x),
    \end{equation}
    where $T_k(x)$ denotes the $k$-th Chebychev polynomial of the first kind and $d=\mathcal{O}(\ceil{1/\pi\varepsilon})$.
    The query complexity in $U$ is $\mathcal{O}(d)$ and the overall circuit depth is $\mathcal{O}(n d)$ with $\mathcal{O}(n)$ needed ancillas. The classical overhead cost for computing the circuit is $\mathcal{O}(\polylog(2^n/\varepsilon^2))$.
\end{lemma}
\begin{proof}
    Based of the proof of the approximation accuracy of $P$ (outsourced to the Appendix in \Cref{lem:absvalapprox}), we can use \Cref{thm:amplitude-arithmetic} to yield the stated complexities.
\end{proof}

\subsection{Assembling QSim}\label{subsec:meth:data-readout}
In this section, we present an approach to solve SLEs of the form $A_x \vec{y} = \vec{b}_x$ using the QSVT in practice. Further, we show how to extract specific summary statistic information from the resulting SLE result, to yield a concrete implementation of the QSim operator specified in \Cref{thm:quso}.

\begin{corollary}[Quantum SLE solver for $A_x \vec{y} = \vec{b}_x$]\label{lem:QSLE-with-x-dep}
    Given a system of linear equations of the form $A_x \vec{y} = \vec{b}_x$ with $s_r$ denoting the minimal row- and $s_c$ the minimal-column-sparsity of $A_x$ over all $x$ with entries of the form $a_{ij}x_{k(i,j)}$ where $k:\left[n\right]^2\rightarrow\left[m\right]$ maps every entry index of $A$ to an index of $x$, as well as an upper bound $\sigma_{\max}^{*}$ on the maximal singular value of all $A_x$, a lower bound $\sigma_{\min}^{*}>0$ on the minimal non-zero singular values of all $A_x$, and $b_x$ with entries of the form $b_{i}x_{k(i)}$ for all $x$, then we can compute a rescaled version of $\vec{y}=A_x^+ \vec{b_x}$ in a circuit of depth $\mathcal{\tilde{O}}(\polylog(N)\kappa^{*}\sqrt{s_rs_c})$ using $\mathcal{\tilde{O}}(N^2)$ ancillary qubits.
\end{corollary}

\begin{lemma}[Rescaling the quantum SLE solvers output]\label{lem:rescalingSLEresult}
    Given a quantum linear systems solver that aims to compute $\ket{y}=A_x^+ \vec{b}/||A_x^+\vec{b}\,||_2$, i.e., a rescaled version of the actual result $\vec{y}=A^+ \vec{b}$. As evident by \Cref{rem:inputproblem}, the subnormalization factor induced by the block-encoding of $A_x$ is of order $\mathcal{O}(\sigma_{\max}^{*}\sqrt{s_rs_c})$. We can then use Quantum Amplitude Estimation as described in \Cref{thm:QAE} on the state of the ancillas being in the $\ket{0}$ state to compute the actual value of $\sigma_{\max}^{*}\sqrt{s_rs_c}\left|\left|A^+ \ket{b}\right|\right|_2/2$, i.e., the amplitude of the state $\ket{0}^{\otimes a}\ket{x}$. The basis-encoded two's complement representation of this amplitude can then be used to properly rescale computations involving $\ket{x}$. Note that for calculating $||A^+\vec{b}\,||_2$ from $\left|\left|A^+ \ket{b}\right|\right|_2$ we can use the equality $||A^+\vec{b}\,||_2 = \left|\left|A^+ \ket{b}\right|\right|_2 ||\vec{b}\,||_2$ and the fact that $||\vec{b}\,||_2$, $\sigma_{\max}^{*}$ and $\sqrt{s_rs_c}$ are necessarily known.
\end{lemma}

\begin{theorem}[Assembling QSim]\label{thm:qsim}
    We can implement a quantum circuit computing $\textnormal{QSim}\ket{x}\ket{0}^{\otimes m} = \ket{x}\ket{u_j(s(x))}$ for any $j\in\left[3\right]$ and $s:x\mapsto A_x^+ b_x$ and $u_1:y\mapsto y_i$, $u_2:y\mapsto \left\langle y, z\right\rangle$, $u_3:y\mapsto y^\top H y$ for a given state preparation of $z$ and block-encoding of $H$, where $s$ represents an SLE of which we are given what is also assumed in \Cref{lem:QSLE-with-x-dep} in a circuit of depth $\mathcal{\tilde{O}}(\polylog(N)\kappa^{*}\sqrt{s_rs_c}/\varepsilon\delta)$ using $\mathcal{\tilde{O}}(N^2)$ ancillary qubits where $\varepsilon$ denotes the QAE accuracy and $0<\delta<1$ the success probability of QAE. An example circuit for $u_1$ is shown in \Cref{fig:quso-U_f}.
\end{theorem}
\begin{proof}
    We use \Cref{lem:QSLE-with-x-dep} to implement the SLE solver and \Cref{thm:QAE} for $u_1$, \Cref{cor:fid} for $u_2$, and \Cref{lem:exp-val} for $u_3$ to compute the rescaled version of $\ket{u(s(x))}$. To fix the scaling, we run \Cref{lem:rescalingSLEresult} with a separate ancillary register to compute the rescaling factor, which is then used in a quantum arithmetic step to compute $\ket{u(s(x))}$.
\end{proof}

\begin{figure*}[htbp]
\centering
\input{tikz/quso-U_f}
\caption{Quantum circuit implementing QSim from \Cref{fig:quso} when the summary statistic information of interest is stored in a single amplitude (cf. \Cref{thm:QAE}). $\textnormal{QLSA}(A_x,b_x)$ denotes a quantum linear system algorithm for the SLE $A_x \vec{y} = \vec{b}_x$ (cf. \Cref{cor:QLS}). The\begin{quantikz}\zctrl{.}\end{quantikz}symbol denotes a controlled $Z$-gate with the control of a subset of wires actuating upon $\ket{0}$ instead of $\ket{1}$. The\begin{quantikz}\pctrl{}\end{quantikz}symbol denotes specific control operators representing the controls in \Cref{fig:QDAC-bitstrings} although the required QDAC is implemented via the circuit in \Cref{fig:QDAC-Wx}.}
\label{fig:quso-U_f}
\end{figure*}

\section{Examples}
\label{sec:examples}
In this section, we show how two industrial applications (one from the energy sector and one from structural engineering) can be formulated as QuSO problems and how our methodology allows up to exponential quantum speedups to solve them.

\subsection{Unit Commitment}
As an example for an SLE with the structure $A \vec{y}=\vec{b}_x$, and an application of \Cref{thm:QDAC-bitstrings} (and \Cref{cor:stateprep_of_arbitrary_bx} respectively), we now show how a highly relevant problem from industry (i.e., the unit commitment problem) can be formulated as a LinQuSO problem and identify that it can be solved via \Cref{thm:quso} and \Cref{thm:qsim}. The unit commitment problem is a MINLP-type optimization problem that concerns the amount of power that is generated at each power generator in a power grid to provide enough power for an estimated demand at the lowest cost. The simulation problem that contributes to the costs is called the power flow problem, which concerns the calculation of the amount of power flowing through every transmission line $\vec{\rho}$, given the power in- and outputs at each node (also called bus) in the power grid $p_i$. The cost that typically needs to be extracted from the result of the power flow problem is of the form $|\langle \vec{c}, \vec{\rho}_x\rangle|$, where $\vec{c}$ denotes the linear cost factor to transmit power over each transmission line. While the unit commitment problem has many other cost factors and constraints (that could be incorporated into our QuSO solver as described in \Cref{rem:MILPQuSO,rem:contquso,rem:contquso}), we will focus on a reduced form of it which only considers the power flow and the costs resulting from it. More specifically, we investigate a specific but very common linear approximation of AC power flow, the DC power flow approximation.

To simplify the definition of the exact form of the unit commitment problem that we want to investigate, we now introduce some notation together with a well-known result from graph theory.

\begin{lemma}\label{lem:cofactors-are-invertible}
    Given the Laplacian matrix $\mathcal{L}$ of an $N$-node graph $\mathcal{G}$, every cofactor $\Lbarc\coloneqq\mathcal{L}\left[i\right]$ (i.e., the matrix that results when the $i$-th row and $i$-th column are removed from $\mathcal{L}$, where $i\in\left[N\right]$) is invertible.
\end{lemma}
\begin{proof}
    By the matrix tree theorem~\cite{doi:10.1073/pnas.40.10.1004}, the determinant of all cofactors is equal to the number of spanning trees of $\mathcal{G}$, and by Kirchhoff's theorem~\cite{1086426}, this number is equal to the product of all non-zero eigenvalues of $\mathcal{L}$ divided by $N$. 
\end{proof}

\begin{definition}[Power flow focused unit commitment]\label{def:UCP}
    Given a power grid in form of a graph $\mathcal{G}=(V,E)$ of its transmission lines weighted by their susceptances $\vec{b}_{ij}\in\mathbb{R}$, $V=\left\lbrace 1,...,N\right\rbrace$, a fixed value how much power $\vec{p}_i\in\mathbb{R}$ is generated or consumed (in this case, $\vec{p}_i\in\mathbb{R}^-$) by each generator $i\in G=\left\lbrace 1,...,M\right\rbrace\subset V$ and load $i\in L \coloneqq V\setminus G$, a reference bus $r\in V$ (wlog. $r=N$), as well as fixed values for the linear cost factor of using each transmission line $\vec{c}_{ij}\in\mathbb{R}$, then we define the problem of power flow focused unit commitment by $\argmin_x \langle \vec{c}, |\vec{\rho}_x|\rangle$ s.t. $\sum_i \vec{p}_x \approx 0$, i.e., optimizing which generators $G_x\coloneqq \left\lbrace i \in G : x_i = 1\right\rbrace$ should be operating to minimize the overall power transmission costs using the decision variables $x\in\left\lbrace 0,1\right\rbrace^M$ while the net power input is roughly equal to the net power output. Here, $\vec{\rho}_x\coloneqq B'\Pi\Bbarc^{-1}\vec{\pbar}_x$ denotes the power flowing over all transmission lines, where $\vec{\pbar}_x$ is the reduced form of the power in-/output vector $\vec{p}_x$ (i.e., $\vec{p}_x$ without its $r$-th entry), which is defined as either $\vec{p}_i$ iff $i\in L$ or $x_i \vec{p}_i$ iff $i\in G$, further, $\mathcal{B}$ is defined as the Laplacian matrix of $\mathcal{G}$ with weights $\vec{b}_{ij}$ and $\Bbarc\coloneqq\mathcal{B}\left[r\right]$ (cf. \Cref{lem:cofactors-are-invertible}). Finally, $\Pi$ denotes the projection matrix mapping $\vec{\theta}\coloneqq\Bbarc^{-1}\vec{\pbar}_x$ onto $\vec{\theta}'\coloneqq(\vec{\theta}_1,...,\vec{\theta}_1,\vec{\theta}_2, ..., \vec{\theta}_2,...,\vec{\theta}_N,...,\vec{\theta}_N)$ where each $\vec{\theta}_i$ appears exactly $d_i\coloneqq\left|\left\lbrace j\in V : (i,j) \in E\right\rbrace\right|$ often and $\vec{\theta}_r=0$ for the chosen reference bus. Lastly, $B'$ takes the form of an $|E|\times |E|$ dimensional matrix with diagonal entries $(\vec{b}_{1 e(1,1)},...,\vec{b}_{1 e(1,d_i)},...,\vec{b}_{N e(N,1)},...,\vec{b}_{N e(N,d_N)})$, where a given function $e:V\times \mathbb{N}\rightarrow V$ yields the $j$-th neighbor of node $i$, and off-diagonal entries $-\vec{b}_{i e(i, k)}$ for the $(k +\sum_{\iota=1}^{i-1} \iota d_\iota)$-th row.
\end{definition}

By $s(x)\coloneqq\vec{\theta}=\Bbarc^{-1}\vec{\pbar}_x$ and $u(s(x))\coloneqq\langle \vec{c}, |B'\Pi s(x)|\rangle$ it becomes obvious, that \Cref{def:UCP} describes a LinQuSO problem as defined in \Cref{def:LinQuSO}. As this form of summary statistic information extraction is already covered by \Cref{cor:fid} and \Cref{lem:absval} up to $B'$ and $\Pi$, the concatenation of these results can be used to implement a quantum algorithm for solving this power flow focused unit commitment problem. For the implementation of $B'\Pi$ in a quantum circuit, we can do the following: Let $d\coloneqq \max_{i\in\left[N\right]} \lceil \log_2 d_i\rceil$, then $\ket{\theta''}\coloneqq\ket{\theta}\ket{+}^{\otimes d}$ yields a quantum version of $\vec{\theta}'$, where $\ket{\theta''}$ is defined with the same basic structure of $\vec{\theta}'$, but with each $\theta_i$ being repeated exactly $d$ times while also introducing the normalization factor $1/\sqrt{2^d}$ for each entry. To match this enlarged version of $\vec{\theta}'$, we have to adapt $B'$ accordingly, i.e., adding $0$-rows and -columns for each entry where $\ket{\theta''}$ exceeds $\vec{\theta}'$. The matrix resulting from this enlargement $B''$ can then be block-encoded via \Cref{thm:OCOBE} analog to the block-encoding of $\Bbarc$ -- the only differences being the reduced sparsity (i.e., two) and the increased dimensionality (i.e., $Nd$). This block-encoding introduces a subnormalization factor of $4\max_{ij}|\vec{b}_{ij}|$, which has to be taken into consideration as a rescaling factor of the result of the scalar product $\langle \vec{c},\vec{\rho}_x\rangle$ (i.e., a rescaling factor on top of the mandatory rescaling discussed in \Cref{lem:rescalingSLEresult}). Note that the state preparation of $\vec{c}$ also has to take the empty rows in $\ket{\theta''}$ into consideration (which the protocol of \Cref{thm:CQSP} easily allows for). Overall the proposed implementation of $B'\Pi$ only affects the computational complexity effectively by a constant factor as the number of ancillary qubits only increases by $\mathcal{O}(N 2^d)=\mathcal{O}(N d_{\max})$ and the QAE runtime only increases linearly with the, also basically constant, rescaling factor of $4\max_{ij}|\vec{b}_{ij}|$.

Based off \Cref{thm:qsim}, the runtime of this algorithm is predominantly dependent on the condition number, as a power grid is generally quite sparse. In the following, we show how graph properties like the maximum node degree and the graph's conductance can be used to bound the condition number. Further, we show that while the condition number often scales linearly wrt. the number of nodes in real-world power grids (cf.~\cite{pareek2024demystifyingquantumpowerflow}), any power grid that takes the form of a so-called expander graph has a constant condition number. In the following we focus on the eigenvalues of $\Bbarc$ instead of the singular values, which is sufficient, as $\sigma_i = |\lambda_i|=\lambda_i$ for Laplacian matrices, because Laplacian matrices are always normal, and as all susceptances can be assumed to be positive (cf. Ref.~\cite{Chethan2024})~\cite[Thm.~4.2]{Mohar1991}).

\begin{lemma}[Gershgorin circle theorem~\protect{\cite{zbMATH02560682}}]\label{lem:gershgorin}
    Every eigenvalue of $A\in \mathbb{C}^{n \times n}$ lies in at least one Gershgorin disc $D_i\coloneqq \left\lbrace x\in\mathbb{C} : |a_{ii} + x| \leq r_i \right\rbrace$, where $r_i\coloneqq \sum_{j\neq i} |a_{ij}|$.
\end{lemma}

\begin{corollary}\label{cor:GershgorinUpperBound}
     For a graph $\mathcal{G}=(V,E)$ with edge weights $(a_{ij})_{(i,j)\in V^2}\in \mathbb{R}$ and its Laplacian matrix $\mathcal{L}$, the maximal eigenvalue $\lambda_{\max}$ of every reduced Laplacian matrix $\Lbarc$ is bound from above by $2 d_{\max}$, where $d_{\max} \coloneqq \max_{i\in V} d_i$ denotes its maximum degree and $d_i\coloneqq\sum_{(i,j)\in E} |a_{ij}|$.
\end{corollary}
\begin{proof}
    This is a consequence of all eigenvalues of $\Lbarc$ lying within the Gershgorin discs of $\mathcal{L}$ by the Cauchy interlacing theorem~\cite{CauchyInterlacing}, combined with $a_{ii}=\sum_{(i,j)\in E} a_{ij}$ and $r_i=\sum_{(i,j)\in E} |a_{ij}|$, s.t. $\lambda_{\max}\leq 2 d_{\max}$.
\end{proof}

Given these insights from graph theory, we can see that the largest eigenvalue of $\Bbarc$ is typically reasonably small, as the maximal degree in sparse graphs hardly scales with an increasing number of nodes. This observation is also mirrored in real-world power grids (cf.~\cite{pareek2024demystifyingquantumpowerflow}). Conversely, the smallest eigenvalue $\lambda_2$ can get up to quadratically small wrt. the number of nodes as $4/N\textnormal{diam}(G)\leq \lambda_2$ as long as all susceptances are positive (which is typically the case in practice~\cite{Chethan2024})~\cite[Thm.~4.2]{Mohar1991}. Aiming for tighter bounds we now discuss the well-known Cheeger inequality, which will be the key for identifying the maximum possible quantum speedup for solving the power flow simulation problem as defined in \Cref{def:UCP}.

\begin{definition}[Fiedler value]
    We call the second smallest eigenvalue of a given Laplacian matrix the Fiedler value (also known as the algebraic connectivity).
\end{definition}

\begin{lemma}[Cheeger's inequality~\protect{\cite{Cheeger}}]\label{lem:CheegerIneq}
    Given a connected graph $\mathcal{G}=(V,E)$ with edge weights $(a_{ij})_{(i,j)\in V^2}\in \mathbb{R}^+$ and its Laplacian matrix $\mathcal{L}$, then the conductance of $\mathcal{G}$
    \begin{equation}
        \varphi(\mathcal{G})\coloneqq\min_{\substack{S\subset V \\ S \neq \varnothing}} \dfrac{\left|\partial S\right|}{\min\left(\textnormal{vol}(S),\textnormal{vol}(\Bar{S})\right)}
    \end{equation}
    gives lower and upper bounds to $\mathfrak{L}$'s Fiedler value $\nu_{2}$ via
    \begin{equation}
        \dfrac{\varphi(\mathcal{G})^2}{2}\leq \nu_{2}\leq 2 \varphi(\mathcal{G}),
    \end{equation}
    where $\mathfrak{L}\coloneqq D^{-1/2}\mathcal{L}D^{-1/2}$ denotes the normalized Laplacian matrix with $D\coloneqq\textnormal{diag}(d_1,...,d_N)$,
    $\left|\partial S\right|\coloneqq \sum_{i\in S}\sum_{j\in \Bar{S}}a_{ij}$ denoting the weight of the edges connecting $S$ with $\Bar{S}\coloneqq V\setminus S$, and $\textnormal{vol}(S)\coloneqq\sum_{i\in S}\sum_{j\in V} a_{ij}$ called the volume of $S$.
\end{lemma}

\begin{lemma}[Relating Fiedler values of $\mathcal{L}$ and $\mathfrak{L}$]\label{lem:FiedlerRelations}
    Given an undirected $N$-node graph $\mathcal{G}=(V,E)$ with edge weights $(a_{ij})_{(i,j)\in V^2}\in \mathbb{R}^+$, its Laplacian matrix $\mathcal{L}$ with associated Fiedler value $\lambda_2$ and its normalized Laplacian matrix $\mathfrak{L}$ with associated Fiedler value $\nu_2$, then:
    \begin{equation}
        d_{\min}\nu_2 \leq \lambda_2 \leq d_{\max} \nu_2.
    \end{equation}
\end{lemma}
\begin{proof}
    We only show the proof for the first inequality, as both proofs are highly analogous. First, we note that all considered matrices here are symmetric, allowing us to form an orthonormal basis of $\mathcal{L}$ based on the spectral theorem~\cite{spectraltheorem}. Due to the structure of Laplacian matrices' rows summing to zero (note that $\mathfrak{L}$ also has the same structure), the smallest eigenvalue of $\mathcal{L}$ as well as smallest eigenvalue of $\mathfrak{L}$ is always $0$, and their associated eigenvector from a respective orthonormal basis is $1_N\coloneqq 1/\sqrt{N} (1,...,1)\in\mathbb{R}^N$. Then choose $w$ as the orthonormal basis vector associated with the second smallest eigenvalue of $\mathcal{L}$, i.e., $\mathcal{L}w=\lambda_2 w$ with $w^{\top}w=1$ and $\langle w,1_N\rangle = 0$. Then we can bound $\nu_2$ via
    \begin{align*}
        \nu_2 = \min_{\substack{\langle v,1_N\rangle = 0 \\ v \neq 0}} \dfrac{v^\top \mathfrak{L}v}{v^\top v} \leq \dfrac{w^\top \mathcal{L} w}{w^\top D w}=\dfrac{\lambda_2}{\sum_i d_i w_i^2}\leq \dfrac{\lambda_2}{d_{\min}},
    \end{align*}
    where the choice of $v\coloneqq D^{-1/2}w$ is valid as (1) clearly $D^{-1/2}w\neq 0$ due to $w\neq 0$, and (2) $\langle D^{-1/2}w,1_N\rangle = 0$ by $\langle D^{-1/2}w,1_N\rangle = \sum_i w_i/\sqrt{d_i}\leq \langle w,1_N\rangle/d_{\max} = 0$ and $\langle D^{-1/2}w,1_N\rangle = \sum_i w_i/\sqrt{d_i}\geq \langle w,1_N\rangle/d_{\min}  = 0$.
\end{proof}

\begin{lemma}[Lower bound for $\Lbarc$'s $\lambda_{\min}$]\label{lem:boundingLaplacianCofactorsSmallestEigenvalue}
    Given an undirected $N$-node graph $\mathcal{G}=(V,E)$ with edge weights $(a_{ij})_{(i,j)\in V^2}\in \mathbb{R}^+$, its Laplacian matrix $\mathcal{L}$ with associated Fiedler value $\lambda_2$ and its normalized Laplacian matrix $\mathfrak{L}$ with associated Fiedler value $\nu_2$, then we can provide the following lower bound on $\Lbarc$'s smallest eigenvalue $\lambda_{\min}$:
    \begin{equation}
        \dfrac{d_{\min}\varphi(\mathcal{G})^2}{2}\leq  d_{\min}\nu_2 \leq \lambda_2 \leq \lambda_{\min}.
    \end{equation}
\end{lemma}
\begin{proof}
    The first inequality is a form of \Cref{lem:CheegerIneq}, the second inequality is the result of \Cref{lem:FiedlerRelations}, and the last inequality follows from the Cauchy interlacing theorem~\cite{CauchyInterlacing}, which implies that all eigenvalues of $\Lbarc$ lie within the range of eigenvalues of $\mathcal{L}$.
\end{proof}

Having established a lower bound for the smallest eigenvalue of $\Bbarc$ through \Cref{lem:boundingLaplacianCofactorsSmallestEigenvalue} by means of the conductance of the underlying power grid susceptances, we now show that there exists a vast number of graphs with constant condition numbers for $\Bbarc$, i.e., so-called expander graphs and their optimal representatives: Ramanujan graphs.

\begin{definition}[Expander graphs]\label{def:expander}
    We call a connected graph $\mathcal{G}$ a $(d,\varepsilon)$-expander iff $d_{\max}\leq d$ and $0<\varepsilon\leq \varphi(\mathcal{G})$.
\end{definition}

\begin{definition}[Ramanujan graphs]\label{def:ramanujan}
    We call a connected, $d$-regular, $N$-node graph $\mathcal{G}=(V,E)$ Ramanujan, iff $\lambda(\mathcal{G})\coloneqq \max_{i\neq 1}|\lambda_i|\leq 2\sqrt{d-1}$, where $\lambda_1\geq \lambda_2 \geq ... \geq \lambda_N$ denote the eigenvalues of $\mathcal{G}$'s adjacency matrix $A$. Note that $\lambda_1=d$ for $d$-regular graphs, as every row in $A$ sums to $d$ such that $A1_N=d1_N$.
\end{definition}

\begin{lemma}[Ramanujan graphs are expanders]\label{lem:ramanujanexpanders}
    Every $N$-node $d$-regular Ramanujan graph is a $(d,\varepsilon)$-expander with $0<\varepsilon\coloneqq 1/2 - \sqrt{d-1}/d$ for $d\geq 3$.
\end{lemma}
\begin{proof}
    Combining \Cref{lem:FiedlerRelations} with \Cref{lem:CheegerIneq} for $d_{\min}=d$ yields $\nu_2 = \lambda_2 /2d \leq \varphi(\mathcal{G})$. Let $w$ be an eigenvector of $\mathcal{L}$ that corresponds to $\lambda_2$ (i.e., $\mathcal{L} w = \lambda_2 w$), then $\mathcal{L}w = (d \mathbb{I} -A)w$ implies that $w$ is also an eigenvector of $A$ with corresponding eigenvalue $\lambda(A)\coloneqq d-\lambda_2$, as $Aw=\left(d-\lambda_2\right)w$. As $\mathcal{G}$ is connected, we have that $\lambda_2>0$, implying $\lambda(A)=(d-\lambda_2)\neq d$, so that we can bound $\lambda(A)\leq \lambda(\mathcal{G})$ completing the proof for the following bounds:
    \begin{equation*}
        \dfrac{1}{2}-\dfrac{\sqrt{d-1}}{d}\leq\dfrac{d-\lambda(\mathcal{G})}{2d}\leq \dfrac{d-\lambda(A)}{2d}=\dfrac{\lambda_2}{2d}\leq \varphi(\mathcal{G}).
    \end{equation*}
\end{proof}

\begin{lemma}[Condition number of Ramanujan graphs]\label{lem:Ramanujankappa}
    The condition number of each cofactor of the Laplacian of an $N$-node $d$-regular Ramanujan graph $\Lbarc$ can be bounded from above by a constant for all $d\geq 3$.
\end{lemma}
\begin{proof}
    $\lambda_{\max}\leq \lambda_{\max}^{*}\coloneqq 2d$ is the result of \Cref{cor:GershgorinUpperBound}. Further, \Cref{lem:ramanujanexpanders,lem:boundingLaplacianCofactorsSmallestEigenvalue} yield
    \begin{equation*}
        \lambda_{\min}^{*}\coloneqq \dfrac{d \left(\frac{1}{2}-\frac{\sqrt{d-1}}{d}\right)^2}{2}\leq \dfrac{d\varphi(\mathcal{G})^2}{2} \leq \lambda_{\min},
    \end{equation*}
    which implies the stated result via
    \begin{equation*}
        \kappa^{*}\coloneqq\dfrac{\lambda_{\max}^{*}}{\lambda_{\min}^{*}}=\dfrac{2d}{\frac{d}{2} \left(\frac{1}{2}-\frac{\sqrt{d-1}}{d}\right)^2}=\dfrac{4}{ \left(\frac{1}{2}-\frac{\sqrt{d-1}}{d}\right)^2}=\mathcal{O}(1).
    \end{equation*}
    
\end{proof}

\begin{remark}[Abundance of expander graphs]
    While it has not yet been shown whether or not infinitely many $d$-regular Ramanujan graphs exist for arbitrary $d$, we know that there exist infinitely many $d$-regular Ramanujan graphs for some more specific $d$ (e.g., $d-1$ being a prime power~\cite{MORGENSTERN199444}). When relaxing the property of being Ramanujan to weakly-Ramanujan however, i.e., $\lambda(\mathcal{G})<2\sqrt{d-1} + \epsilon$ for some $\epsilon>0$, even widely common random $d$-regular graphs have this property with high probability~\cite{10.1145/780542.780646}. As such weakly-Ramanujan graphs are also expander graphs by a straightforward extension of \Cref{lem:ramanujanexpanders,lem:Ramanujankappa}, this in principle displays a large class of potential power grid topologies that would be eligible for an exponential quantum speedup. However, due to the very high connectivity required in expander graphs (e.g., the diameter scaling merely logarithmically wrt. the number of nodes~\cite{chung1989diameters}), it is questionable if power grids can reasonably be structured accordingly in practice. Even if power grids could take the form of expanders, it is contestable if currently existing power grids could be adapted accordingly, or if this would only be a consideration for the architecture of newly assembled power grids. Nevertheless, some literature already exists that motivates a condition-number-minimizing design of power grids, see, e.g., Ref.~\cite{5618969}.
\end{remark}



\subsection{Topology Optimization}
Designing mechanical structures requires to find the optimal balance between functional factors like stiffness or aerodynamic properties and constraining factors such as cost, weight or space. Algorithmic approaches show potential in automatically exploring the solution space but are often limited by the costly evaluations of functional properties which requires complex numerical simulations or even experiments. Due to constraints in time and cost, only a few points in the design space can be evaluated and thus a global optimum is often not reachable. An improvement can either be obtained by speeding up the simulation and thus enabling more evaluations or finding a better optimization strategy that more efficiently explores the design space. While we focus on the former, the later will have to be target of investigation for future (numerical) experiments. A typical use case in the domain of topology optimization is finding the maximally stiff structure for a beam under a specific load and weight constraint (cf. \cite{bendsoe2013topology,zie00}). For simplicity, we limit the following description to a system comprised of only one single type of element organized in a uniform mesh. Note however, that such a discretization is often insufficient in problems of industrial relevance to yield accurate results for the simulated physical properties. Nevertheless, it is an important and meaningful step towards exploring real-world QuSO problems.

\begin{definition}[Topology optimization by distribution of isotropic material]\label{def:TOP}
By discretizing the domain $\Omega \coloneqq \left[0,\omega_x\right]\times \left[0,\omega_y\right]\subset \mathbb{R}^2$ that shall contain the body of the mechanical structure into an $n_x\times n_y$ grid of uniform linear right-triangle-shaped finite elements as shown in \Cref{fig:TOP}, where $\omega_x,\omega_y,n_x,n_y\in\mathbb{N}$, a corresponding topology optimization problem can be mathematically described by
\begin{equation}\label{eq:topology-cost}
   \argmin_{\substack{x\in X\subset \lbrace 0,1\rbrace ^{2 n_x n_y} \\ \sum_{i}x_i / 2n_xn_y =v}} \vec{u}^\top K_{x}\vec{u},
\end{equation}
where the decision variables $x$ denote whether the corresponding finite element with index $i$ should be filled with material (i.e., $x_i=1$) or not (i.e., $x_i = 0$), and $0<v\leq 1$ is a given maximum fraction of volume to be filled with material. $X$ denotes the set of physically realizable shapes, i.e., (1) each filled triangle is connected to another filled triangle by sharing two nodes, (2) the nodes on which external force is applied are part of a filled triangle and (3), the given fixed points are also part of a filled triangle. $(K_x)_{jk}\coloneqq \sum_i (K_x^{(i)})_{\pi^{-1}(j,k)}$ denotes the entries of the global stiffness matrix which is assembled through the local stiffness matrices of each element $K_x^{(i)}\coloneqq \Delta \left(B^{(i)}\right)^{\top} D^{(i)}_x B^{(i)}$ and a projector $\pi$ mapping the indices of the local degrees of freedom in the local stiffness matrices to the corresponding indices of the global degrees of freedom given some numbering of all nodes in the mesh. Here, $\Delta\coloneqq \omega_x\omega_y/2n_xn_y$ denotes the area of each finite element, and $B^{(i)}$ denotes the strain-displacement matrix of the $i$-th finite element whose vertices $(x^{(i)}_1,y^{(i)}_1)\in\Omega$, $(x^{(i)}_2,y^{(i)}_2)\in\Omega$, and $(x^{(i)}_3,y^{(i)}_3)\in\Omega$ are enumerated counterclockwise, s.t.
\begin{equation}
    B^{(i)}\coloneqq \frac{1}{2\Delta}\begin{pmatrix}
\beta^{(i)}_1 & 0 & \beta^{(i)}_2 & 0 & \beta^{(i)}_3 & 0 \\
0 & \gamma^{(i)}_1 & 0 & \gamma^{(i)}_2 & 0 & \gamma^{(i)}_3 \\
\gamma^{(i)}_1 & \beta^{(i)}_1 & \gamma^{(i)}_2 & \beta^{(i)}_2 & \gamma^{(i)}_3 & \beta^{(i)}_3
\end{pmatrix},
\end{equation}
where $\beta^{(i)}_1 \coloneqq y_2^{(i)} - y_3^{(i)}$, $\beta^{(i)}_2 \coloneqq y_3^{(i)} - y_1^{(i)}$, $\beta^{(i)}_3 \coloneqq y_1^{(i)} - y_2^{(i)}$, $\gamma^{(i)}_1 \coloneqq x_3^{(i)} - x_2^{(i)}$, $\gamma^{(i)}_2 \coloneqq x_1^{(i)} - x_3^{(i)}$, and $\gamma^{(i)}_3 \coloneqq x_2^{(i)} - x_1^{(i)}$. Further, assuming all filled elements are made of identical isotropic material, the plane strain matrix $D^{(i)}_x$ is defined as
\begin{equation}
    D^{(i)}_x\coloneqq \frac{E(x_i)}{(1+\nu)(1-2\nu)}\begin{pmatrix}
        1 - \nu & \nu & 0 \\
        \nu & 1 - \nu & 0 \\
        0 & 0 & \frac{1 - 2\nu}{2}
    \end{pmatrix},
\end{equation}
where $\nu$ denotes the Poisson's ratio (e.g., $\nu=0.3$ for typical structure steel), and $E(x_i)\coloneqq E_{\min} + x_i \left(E_0 - E_{\min}\right)$ the Young's modulus of each element, depending on whether it is filled (i.e., $E(x_i)=E_0$, where, e.g., $E_0=210000 \textnormal{MPa}$ for typical structure steel) or not (i.e., $E(x_i)=E_{\min}>0$). Finally, $\vec{u}$ denotes the vector of displacements for the degrees of freedom in the nodes of the structure of the finite element mesh, which is implicitly given through the system of equations $K_x \vec{u}= \vec{f}$, where $\vec{f}$ is the vector of external forces acting on the structure for each degree of freedom. Note that due to the degrees of freedom in fixed nodes being fixed, we have to delete the rows and columns from $K_x$ and $\vec{u}$ that concern these degrees of freedom in all places where they appear in the definitions above (which we did not consider in this to enhance readability).
\end{definition}

\begin{figure}[hbtp]
\centering
\begin{tikzpicture}[scale=0.75]  
    \tikzset{every path/.style={thick}}

    \newcounter{trianglenumber}
    \setcounter{trianglenumber}{1}
    
    \foreach \x in {0, 2, 4, 6} {
        \foreach \y in {0, -2, -4} {
            \draw (\x, \y) rectangle (\x+2, \y-2);
            \draw (\x+2, \y) -- (\x, \y-2);
            
            \node at (\x+0.5, \y-0.5) {\thetrianglenumber};
            \stepcounter{trianglenumber} 
            
            \node at (\x+1.5, \y-1.5) {\thetrianglenumber};
            \stepcounter{trianglenumber} 
        }
    }
    
    \foreach \x in {0, 2, 4, 6, 8} {
        \foreach \y in {0, -2, -4, -6} {
            \filldraw (\x, \y) circle (2pt); 
        }
    }
    
    \foreach \y in {0, -2, -4, -6} {
        \draw (-0.4, \y-0.4) -- (0, \y) -- (-0.4, \y+0.4) -- cycle;
        \draw (-0.4, \y+0.4) -- (-0.55, \y+0.25);  
        \draw (-0.4, \y+0.25) -- (-0.55, \y+0.1);  
        \draw (-0.4, \y+0.1) -- (-0.55, \y-0.05);  
        \draw (-0.4, \y-0.05) -- (-0.55, \y-0.2);  
        \draw (-0.4, \y-0.2) -- (-0.55, \y-0.35);  
    }
    
    \draw[-{Stealth[length=4mm, width=3mm]}] (8, -6) -- (8, -7);

    \draw[{Stealth[length=3mm]}-{Stealth[length=3mm]}] (0, 0.5) -- (8, 0.5);  
    \node at (4, 0.8) {\(\omega_x\)};  

    \draw[{Stealth[length=3mm]}-{Stealth[length=3mm]}] (8.5, 0) -- (8.5, -6);  
    \node at (9, -3) {\(\omega_y\)};  

\end{tikzpicture}
\caption{Visualization of an example to the optimization problem defined in \Cref{def:TOP}, where $n_x = 4$, $n_y = 3$, and where the external force applied $\vec{f}$ only acts upon the vertical degree of freedom of the bottom right node in the mesh. The numbers shown depict possible indices of the finite elements. The triangles to the left of the left-most grid nodes indicate that these nodes are fixed.}
\label{fig:TOP}
\end{figure}

\begin{remark}[Topology optimization is LinQuSO]
    Intuitively, the described optimization problem (cf. Ref.~\cite{Andreassen2011} for a similar formulation) describes the aim at finding a physically meaningful material assignment within a fixed volume fraction $v$ that leads to minimal compliance $c(x)\coloneqq \vec{u}^\top K_x \vec{u}$ of the structure under a given load $\vec{f}$. Even though further constraints and costs (e.g., ease of manufacturing, spatial constraints) could be included, we limit the discussion in this paper to only include the structure's compliance within the cost function as this already forms a LinQuSO problem, i.e., by defining $s(x)\coloneqq K_x^{+} \vec{f}$ and $u(s(x))\coloneqq s(x)^{\top} K_x s(x)$ it becomes clear, that the problem stated in \Cref{def:TOP} is a LinQuSo problem (cf. \Cref{def:LinQuSO}).
\end{remark}

\begin{remark}[Sparsity and condition number estimates]
    As the entries in the global stiffness matrix represent interactions of nodes in the mesh, the number of non-zero entries scales linearly with the maximum amount of neighbors any node can have. By employing the uniform mesh displayed in \Cref{fig:TOP} with a horizontal and vertical degree of freedom for every node, $K_x$ has a sparsity $\leq 26$, as every node's two degrees of freedom are affected by the degrees of freedom of its direct neighbours (there exist at maximum six neighbours with two DOFs in our mesh) and itself. Further, there exists an upper bound on the value of the maximum entry of each local stiffness matrix (due to such bounds being implied by definition for the strain-displacement matrix $B^{(i)}$ and the plane strain matrix $D_x^{(i)}$). Therefore, by the Gershgorin circle theorem \Cref{lem:gershgorin}, we can set a constant upper bound on the maximal singular value of the global stiffness matrix. For a lower bound on the smallest singular value of $K_x$, we resort to the well-known result of the condition number of the global stiffness matrix scaling as $\mathcal{O}(1/h^2)$, where $h$ denotes the size of each finite element~\cite{Yserentant1986}. This implies that the condition number scales linearly in the dimension of the SLE $\mathcal{O}(N)$ and hence the smallest singular value correspondingly can be bound from below with a value $\mathcal{O}(1/N)$. Note however, that by switching from the here employed nodal basis of the finite element space to a hierarchical basis, the condition number could be reduced to $\mathcal{O}(\log N)$, enabling an exponential quantum speedup~\cite{Yserentant1986}. Future work will have to show if our approach can be generalized to the slightly altered decision-variable-dependent SLE that emerges from this different finite element space basis.
\end{remark}

Having established the sparsity and bounds on the largest and smallest singular value, we now show how this problem can be solved with the framework of quantum algorithms proposed in \Cref{sec:methodology}.

By representing the material assignments as a binary vector (where zero represents a void element and one a filled element), the conditional block encoding of $K_x$ can be achieved with the complexity stated in \Cref{thm:OCOBE} by slightly adapting the quantum algorithm presented there to allow for the matrix entries of the type $c + x_k (a_{ij} -c)$ for $c\coloneqq E_{\min}$ constant and $a_{ij}$ denoting the entries of the global stiffness matrix $K_x$ when $x_k=1$ for all $k$. This adaption can be accomplished by analogously using \Cref{lem:FOBE} for block-encoding a matrix with entries $a_{ij}$ while stopping before the AQE step to insert the UCU from \Cref{thm:QDAC-bitstrings} controlled on the matrix index registers to compute the value of the respective $x_{k(i,j)}$ on an ancillary qubit. Then we control the AQE on this ancilla to ensure that the value for $a_{ij}$ is only respected if the corresponding $x_{k(i,j)}$ equals one, additionally we prepare an ancillary register to represent $c$ in basis encoding and add an AQE of this register controlled on the respective decision variable $x_{k(i,j)}$ being zero. After that, we uncompute these steps (except for the AQE) and finally uncompute $O_A$ to conclude the block-encoding. By applying the implementation setup from \Cref{fig:QDAC-Wx} for the UCU, we end up with an algorithm of complexity in the same order of magnitude as \Cref{thm:OCOBE}.

By using \Cref{thm:qsim} with $u_3$ and \Cref{lem:QSP} to input the external force vector $\vec{f}$ we can implement QSim for the stated topology optimization problem. Finally to obtain a result for the topology optimization problem, we can run the QuSO solver defined in \Cref{thm:quso} by intrducing respective penalty terms to ensure the physical meaningfulness of the result (i.e., $x\in X$).

\section{Discussion}
\label{sec:discussion}
In this article, we proposed a novel class of optimization problems, that require summary statistic information on the result of a simulation to compute the cost function or ensure the validity of constraints called Quantum Simulation-based Optimization (QuSO). Further, we introduced an efficient quantum algorithm to perform a specific type of digital to analog conversion to facilitate the combination of the QAOA and the QSVT to efficiently solve QuSO problems. Finally, we exemplified the application of the developed approach to achieve up to exponential quantum speedups for two use cases with potential practical relevance. More precisely, the simulation component of the QuSO problem can be solved in time $\tilde{\mathcal{O}}(\polylog(N)\kappa^{*}s)$ compared to the conjugate gradient method's (i.e., the classical state of the art) $\mathcal{O}(N\kappa s)$ for indefinite $N\times N$ matrices (and $\mathcal{O}(N\kappa s)$ for positive definite matrices) with an upper bound on the condition number $\kappa^{*}\geq \kappa$ and sparsity $s$. In practice, the potential quantum speedup is thus quadratic for condition numbers that scale linearly with the size of the system (e.g., in practical instances for the optimal power flow problem) and exponential if the condition number scales at most logarithmically with the dimension of the SLE (e.g., for the optimal power flow problem on expander graphs). As almost all subroutines used in our quantum solver for QuSO problems require a substantial amount of error-corrected quantum hardware, the first examples for practically relevant quantum speedup will likely require a very small condition number (cf.~\cite{PRXQuantum.2.010103}). While practically relevant use cases with constant or at most logarithmic condition number exist~\cite{Yserentant1986,Bank1988,doi:10.1137/0726080,bramble1991convergence,Yserentant_1993,https://doi.org/10.1002/nme.506}, their practical problem instances can often already be solved within reasonable computational costs using classical algorithms, as these are already quasi-linear in the dimension of the SLE. Therefore, the quadratic quantum speedup to be gained for problems with linear condition number might even have a bigger effect in practice, when quasi-constant-overhead quantum error correction~\cite{8555154} is available in practice.

Once sufficiently large quantum hardware is available to solve real world instances of QuSO problems, the performance of the QAOA against classically employed simulation-based optimization algorithms like Monte-Carlo search, genetic algorithms, response surface methodology, or the SIMP method for topology optimization, should be benchmarked to evaluate if our QuSO solver can also provide practical wallclock-time speedups or even a better solution quality. Based on the fact that typical state-of-the-art optimization algorithms like branch and bound are not necessarily well-suited to solve optimization problems with very large systems of linear equations as constraints, such a benchmark will be particularly interesting, as BnB solvers are typically the state of the art for practically all combinatorial optimization problems that where predominantly the topic of past quantum optimization benchmarks. Another potentially relevant difference to classical state-of-the-art simulation-based optimization solvers, is that our QAOA-driven approach does not necessarily need multiple iterations to find optimal solutions, i.e., measurements from a single quantum circuit can already suffice (assuming predefined or pretrained QAOA parameters, cf.~\cite{sack2021}).

The provided examples demonstrate the ability of the proposed framework of quantum subroutines to solve practically relevant QuSO problems. Stemming on theoretical results from spectral graph theory as well as mesh structure in the finite element method, we exemplified the process of computing the required bounds for the largest and smallest singular values as well as an upper bound on the sparsity. The employed approach for the optimal power flow simulation in context of the unit commitment problem shows the first practically relevant quantum algorithm in the context of recent research in this domain by effectively bypassing the output problem~\cite{9423668,GAO2023110147,10078612,QCPowerFlowQPU,neufeld2024hybrid,Liu2024,pareek2024demystifyingquantumpowerflow}. A logical next step for exploring the applicability of quantum simulation speedups in this context would be investigating the non-linear form of the power flow equations -- cf. Refs.~\cite{freris1968investigation,4111178} for respective classical (but also typically highly approximative) approaches. For the topology optimization use case, we showed a quadratic speedup analogous to the power flow simulation (where speedup is only of order $\mathcal{O}(n^{3/2})$). Further research has to show if our framework of quantum algorithmic subroutines can be applied, or sufficiently extended, to allow for an exponential speedup either using suitable preconditioners~\cite{bramble1990parallel,Bornemann1991,10.1093/imanum/dri037} or, e.g., a hierarchical basis for the mesh~\cite{Yserentant1986}. Notably, the exploration of preconditioners might be especially interesting, as recent literature showed efficient implementations for the respective block-encoding~\cite{quantumFEM}. In general, we strongly expect the existence of more efficient approaches to conduct the block-encoding for simple FEM models, as the matrix entries can easily be described by simple algebraic functions. This could open up QuSO use cases that do not suffer from linear or even quadratic space requirements for the data input, but rather operate within logarithmic space requirements, which would allow for significantly sooner practical applications of the presented approach.

Future work may focus on the identification of other practically relevant QuSO problems. Central questions to decide the applicability of the presented framework are the sparsity of the SLE, bounds on the smallest non-zero and the largest singular values, and optionally any mathematically exploitable structure in $A_x$ and $\vec{b}_x$ for corresponding state preparation oracles. Beyond this, non-linear as well as quantum-native use cases should be investigated accordingly.




\newpage

\begin{acknowledgments}
This paper was partially funded by the German Federal Ministry for Economic Affairs and Climate Action through the funding program ``Quantum Computing -- Applications for the industry'' based on the allowance ``Development of digital technologies'' (contract number: 01MQ22008A). PA acknowledges support from the Munich Quantum Valley, which is supported by the Bavarian state government with funds from the Hightech Agenda Bayern Plus.
\end{acknowledgments}


\bibliography{references}

\appendix
\section{Complexities of subroutines}
The complexities of all algorithms employed for data input, processing and output are displayed in \Cref{tab:qblas}.
\begin{table*}[tbp]
\centering
\begin{tabular}{|c|c|c|c|c|} \thickhline
 \multicolumn{5}{|c|}{Data Input}\\ \thickhline
 operation & qAlgo & depth $+$ query complexity & space & \#ancillas \\ \hline
 {$\!\begin{aligned}
     \mathbb{R}^{N} & \rightarrow \mathcal{H}^{\otimes n} \\
     b & \mapsto \ket{b}
 \end{aligned}$} & QStPr~(\Cref{lem:QSP}) & $\mathcal{O}(\log(N)) + 0$ & $\mathcal{O}(\log{N})$ & $\mathcal{O}(N)$\\\hline
 {$\!\begin{aligned}
     \mathbb{R}^{N} & \rightarrow \mathcal{H}^{\otimes n} \\
     b_x & \mapsto \ket{b_x}
 \end{aligned}$} & QStPr~(\Cref{lem:FCSP}) & $\mathcal{O}(\log(N)) + 0$ & $\mathcal{O}(\log{N})$ & $\mathcal{O}(N^2)$\\\hline
 {$\!\begin{aligned}
     \mathbb{R}^{N\times N} & \rightarrow \textnormal{SU}(n) \\
     A & \mapsto U_A
 \end{aligned}$} & OBE~(\Cref{lem:FOBE}) & $\tilde{\mathcal{O}}(\log(n)) + \mathcal{O}(\log(N))$ & $\mathcal{O}(n+3)$ & $\mathcal{O}(N + \log 1/\varepsilon)$\\\hline
 {$\!\begin{aligned}
     \mathbb{R}^{N\times N} & \rightarrow \textnormal{SU}(n) \\
     A_x & \mapsto U_{A_x}
 \end{aligned}$} & OBE~(\Cref{lem:FCOBE}) & $\tilde{\mathcal{O}}(\log(n)) + \mathcal{O}(\log(N))$ & $\mathcal{O}(n+3)$ & $\mathcal{O}(N^2 + \log 1/\varepsilon)$\\ \thickhline
 \multicolumn{5}{|c|}{Data Processing}\\ \thickhline
 {$\!\begin{aligned}
     \textnormal{SU}(n) & \rightarrow \textnormal{SU}(n) \\
     U_A & \mapsto U_{A^{+}}
 \end{aligned}$} & QSVT~(\Cref{thm:matinv}) & $\mathcal{O}(1) + \mathcal{O}(\kappa\log(\frac{\kappa}{\varepsilon}))$ & $\mathcal{O}(\log{N})$ & $\mathcal{O}(1)$ \\\hline 
 {$\!\begin{aligned}
     \textnormal{SU}(n) & \rightarrow \textnormal{SU}(n) \\
     U_\psi & \mapsto U_{\left|\psi\right|}
 \end{aligned}$} & QSVT (\Cref{lem:absval}) & $\mathcal{O}(n/\varepsilon) + \mathcal{O}(1/\varepsilon)$ & $\mathcal{O}(n)$ & $\mathcal{O}(1/\varepsilon)$\\  \thickhline
 \multicolumn{5}{|c|}{Extracting Summary Statistic Information}\\ \thickhline
   {$\!\begin{aligned}
     \textnormal{SU}(n) & \rightarrow \left[-0.5,0.5\right] \\
     U_{\psi} & \mapsto |\psi_i|
 \end{aligned}$} & QAE (\Cref{thm:QAE}) & $\mathcal{O}(1/\varepsilon\delta) + \mathcal{O}(1/\varepsilon\delta)$  & $\mathcal{O}(n+\log 1/\varepsilon\delta)$ & $\mathcal{O}(1)$  \\ \hline
 {$\!\begin{aligned}
     \textnormal{SU}(n)\times\textnormal{SU}(n) & \rightarrow \left[0,1\right] \\
     U_{\varphi}, U_{\psi} & \mapsto \left|\braket{\varphi | \psi}\right|
 \end{aligned}$} & QAE (\Cref{cor:fid}) & $\mathcal{O}(1/\varepsilon\delta) + \mathcal{O}(1/\varepsilon\delta)$  & $\mathcal{O}(n+\log 1/\varepsilon\delta)$ & $\mathcal{O}(1)$  \\ \hline
  {$\!\begin{aligned}
     \textnormal{SU}(n)\times\textnormal{SU}(n) & \rightarrow \left[-1,1\right] \\
     U_H,U_{\psi} & \mapsto \bra{\psi} H \ket{\psi}
 \end{aligned}$} & QAE (\Cref{lem:exp-val}) & $\mathcal{O}(1/\varepsilon\delta) + \mathcal{O}(1/\varepsilon\delta)$  & $\mathcal{O}(n+\log 1/\varepsilon\delta)$ & $\mathcal{O}(1)$  \\ \hline
\end{tabular}
\caption{Quantum basic linear algebra subroutions (qBLAS) used in this article and their complexities. We use the definition $N\coloneqq 2^n$, where $n\in\mathbb{N}$ and assume that all vectors are normalized and all matrices are real and have a spectral norm lesser or equal than 1.}\label{tab:qblas}
\end{table*}

\section{Implementations of the diagonal block-encoding operators \texorpdfstring{$G_p$}{Gp} and \texorpdfstring{$W_p$}{Wp}}
\label{app:diagonal-blockencoding}

The operators $G_p$ and $W_p$ as defined in Ref.~\cite{rattew2023nonlinear} are displayed in \Cref{fig:diagonal-blockencoding-Gp} and \Cref{fig:diagonal-blockencoding-Wp}.

\begin{figure}[hbtp]
\centering
\input{tikz/diagonal-blockencoding-Gp}
\caption{Quantum circuit implementing the operator $G_p$ as used in the implementation of \Cref{thm:diagonal-blockencoding}.}
\label{fig:diagonal-blockencoding-Gp}
\end{figure}

\begin{figure}[hbtp]
\centering
\input{tikz/diagonal-blockencoding-Wp}
\caption{Quantum circuit implementing the operator $W_p$ as used in the implementation of \Cref{thm:diagonal-blockencoding}.}
\label{fig:diagonal-blockencoding-Wp}
\end{figure}

\section{Polynomial Approximation of the absolute value function}\label{sec:absval}
In the following, we provide a proof for \Cref{lem:absval}, i.e., show that the polynomial stated in \Cref{lem:absval} provides an efficient and accurate approximation of the absolute value function $|\cdot|: [-1,1] \rightarrow [-1,1]$. 

\begin{definition}[Fourier series]\label{def:fourierseries}
The Fourier series of any $2\pi$-periodic function $f \in L^1([0, 2\pi]) $ is defined as
\begin{equation}
    (\mathcal{F} f)(x) \coloneqq \frac{a_0}{2} + \sum_{k=1}^{\infty} (a_k \cos kx + b_k \sin kx)
\end{equation}
via the Fourier coefficients $a_k \coloneqq \frac{1}{\pi} \int_{0}^{2\pi} f(x) \cos kx \, dx$ and $b_k \coloneqq \frac{1}{\pi} \int_{0}^{2\pi} f(x) \sin kx \, dx$ for $k \in \mathbb{N}$ and $a_0 \coloneqq \frac{1}{\pi} \int_{0}^{2\pi} f(x) \, dx$.
\end{definition}

\begin{lemma}\label{lem:absvalfourier}
    The Fourier series of $f(x) \coloneqq |\cos(x)| $ for $x \in \mathbb{R}$ is given by
    \begin{equation}
        (\mathcal{F} f)(x) = \frac{2}{\pi} + \frac{4}{\pi}\sum_{k=1}^{\infty} \frac{(-1)^{k+1}}{4k^2 - 1} \cos(2kx)
    \end{equation}
\end{lemma}
\begin{proof}
As $f(x)=|\cos(x)|$ is a $2\pi$-periodic function in  $L^1([0, 2\pi]$, the Fourier coefficients $a_k $ and $b_k $ for $k \in \mathbb{N} $ can be determined according to the definition stated in \Cref{def:fourierseries}. For the $a_k $ coefficients we thus get
\begin{align*}
    a_k &= \frac{1}{\pi} \int_{0}^{2\pi} f(x) \cos(kx) \, dx \\
    &= \frac{1}{\pi} \int_{0}^{2\pi} |\cos(x)| \cos(kx) \, dx \\
    &= \frac{2}{\pi} \Big( \underbrace{\int_{0}^{\frac{\pi}{2}} \cos(x) \cos(kx) \, dx}_{I_1\coloneqq} - \underbrace{\int_{\frac{\pi}{2}}^{\pi} \cos(x) \cos(kx) \, dx}_{I_2\coloneqq} \Big)
\end{align*}
Using the trigonometric identity $\cos(x) \cos(kx) = \frac{1}{2} \left(\cos((k+1)x) + \cos((k-1)x)\right)$, we can compute $I_1$ as
\begin{equation*}
I_1 = \frac{1}{2} \Big( \underbrace{\int_0^{\frac{\pi}{2}} \cos((k + 1)x) \, dx}_{I_{11}\coloneqq} + \underbrace{\int_0^{\frac{\pi}{2}} \cos((k - 1)x) \, dx}_{I_{12}\coloneqq} \Big)
\end{equation*}
By substituting $u \coloneqq \left(k + 1\right)x$ in $I_{11}$, we get
\begin{align*}
    I_{11} =& \int_0^{\frac{\pi}{2}} \cos(u) \frac{du}{k + 1} = \frac{1}{k + 1} \int_0^{\frac{\pi}{2}} \cos(u) \, du \\
    =& \frac{1}{k + 1} \sin\left((k + 1) \frac{\pi}{2}\right).
\end{align*}
Analogously substituting $v \coloneqq \left(k - 1\right)x$ in $I_{12}$ yields
\begin{align*}
I_{12} =& \int_0^{\frac{\pi}{2}} \cos(v) \frac{dv}{k - 1} = \frac{1}{k - 1} \int_0^{\frac{\pi}{2}} \cos(v) \, dv \\=& \frac{1}{k - 1} \sin\left((k - 1) \frac{\pi}{2}\right).
\end{align*}
Combining these results, $I_1$ takes the form of
\begin{align*}
I_1 =& \frac{1}{2} (I_{11} + I_{12}) \\=& \frac{1}{2} \left( \frac{1}{k + 1} \sin\left((k + 1) \frac{\pi}{2}\right) + \frac{1}{k - 1} \sin\left((k - 1) \frac{\pi}{2}\right) \right).
\end{align*}
As $I_2$ is equivalent to $I_1$ up to the integration interval, we get
\begin{align*}
    I_2 =& \frac{1}{2} \left( \int_{\frac{\pi}{2}}^{\pi} \cos(kx+x) \, dx + \int_{\frac{\pi}{2}}^{\pi} \cos(kx-x) \, dx \right) \\
    =& \frac{1}{2} \left( \tfrac{1}{k+1} \left( \sin((k+1)\pi) - \sin \left( (k+1)\tfrac{\pi}{2} \right) \right)\right) \\ & + \frac{1}{2} \left(\tfrac{1}{k-1} \left( \sin((k-1)\pi) - \sin \left( (k-1)\tfrac{\pi}{2} \right) \right) \right).
\end{align*}
Therefore we can now simplify $a_k$ into
\begin{align*}
    a_k &= \frac{2}{\pi} (I_1 - I_2) \\
    &= \frac{1}{\pi} \left( \frac{2 \sin \left( (k+1)\frac{\pi}{2} \right)}{k+1} + \frac{2 \sin \left( (k-1)\frac{\pi}{2} \right)}{k-1} \right) \\
    &= \frac{2}{\pi} \left( \frac{\sin \left( (k+1)\frac{\pi}{2} \right)}{k+1} + \frac{\sin \left( (k-1)\frac{\pi}{2} \right)}{k-1} \right).
\end{align*}
Further, we recognize that for any odd $k$, i.e., $k = 2m+1$ with $m \in \mathbb{N}$, we have
\begin{align*}
    &\frac{\sin \left( (2m+1+1)\frac{\pi}{2} \right)}{2m+1+1} + \frac{\sin \left( (2m+1-1)\frac{\pi}{2} \right)}{2m+1-1} \\=&  \frac{\sin \left( (2(m+1))\frac{\pi}{2} \right)}{2(m+1)} + \frac{\sin \left( (2m)\frac{\pi}{2} \right)}{2m} = 0,
\end{align*} 
such that we only need to consider even $k$, allowing for the substitution $k \mapsto 2k$ and yielding
\begin{align*}
   a_k &= \frac{2}{\pi} \left( \frac{\sin \left( (2k+1)\frac{\pi}{2} \right)}{2k+1} + \frac{\sin \left( (2k-1)\frac{\pi}{2} \right)}{2k-1} \right) \\
    &= \frac{2}{\pi} \left( \frac{(-1)^k}{2k+1} + \frac{-(-1)^{k}}{2k-1} \right) \\
    &= \frac{2(-1)^k}{\pi} \left( \frac{1}{2k+1} - \frac{1}{2k-1} \right) \\
    &= \frac{2(-1)^k}{\pi} \left( \frac{(2k-1) - (2k+1)}{(2k+1)(2k-1)} \right) \\
    &= \frac{2(-1)^k}{\pi} \cdot \frac{-2}{(2k+1)(2k-1)} \\
    &= \frac{4(-1)^{k+1}}{\pi} \cdot \frac{1}{4k^2 - 1} \\
    &= \frac{4(-1)^{k+1}}{\pi(4k^2 - 1)}.
\end{align*}
Importantly, since $|\cos(x)|$ is even and $\sin(nx) $ is odd, their product is also odd and thus integrates to zero over a symmetric interval (which results by taking the same approach as for $a_k$, i.e., the splitting into $I_1$ and $I_2$), implying that the $b_k$ terms vanish. Thus, the following calculation of $a_0$ completes the proof.
\begin{align*}
a_0 &= \frac{1}{\pi} \int_{0}^{2\pi} |\cos(x)| \, dx \\
&= \frac{2}{\pi} \Big( \int_{0}^{\frac{\pi}{2}} \cos(x) \, dx - \int_{\frac{\pi}{2}}^{\pi} \cos(x) \, dx \Big) \\
&= \frac{2}{\pi} \left(\sin(x) \bigg|_0^{\frac{\pi}{2}} - \sin(x)\bigg|_{\frac{\pi}{2}}^{\pi}\right) \\
&= \frac{4}{\pi}.
\end{align*}
\end{proof}

\begin{lemma}\label{lem:uniformConvergence}
    The Fourier series of $f(x) \coloneqq |\cos(x)| $ uniformly converges towards $(\mathcal{F} f)(x)$ for all $x \in \mathbb{R}$.
\end{lemma}
\begin{proof}
    This a simple application of Dirichlet's Theorem~\cite{Dirichlet1829} stating that Fourier series $(\mathcal{F}f)(x)$ of absolutely integrable $2\pi$-periodic functions that have a finite number of local extrema as well as a finite number of finite discontinuities in each period $f(x)$ converge to $1/2 \left( \lim_{x\rightarrow a^+}f(x) + \lim_{x\rightarrow a^-}f(x)\right)$.
\end{proof}

\begin{corollary}\label{cor:absvalapprox}
    Let $T_n(\cos(x))\coloneqq \cos(nx)$ be defined as the Chebyshev polynomial of the first kind, then the substitution of $x\mapsto\arccos x$ into \Cref{lem:uniformConvergence} yields
    \begin{equation}
        |x|= f_{\infty}(x) \coloneqq \frac{2}{\pi} + \frac{4}{\pi}\sum_{k=1}^{\infty} \frac{(-1)^{k+1}}{4k^2 - 1} T_{2k}(x),
    \end{equation}
     as $T_n (x) = \cos (n\arccos x)$ for $x \in [-1,1]$.
\end{corollary}

\begin{lemma}\label{lem:absvalapprox}
    The following degree-$d$ polynomial is an $\varepsilon$-approximation of $|x|$ for $x \in [-1,1]$ and $\varepsilon=\mathcal{O}(1/d)$
    \begin{equation}
        f_d(x) \coloneqq \frac{2}{\pi} + \frac{4}{\pi}\sum_{k=1}^{d} \frac{(-1)^{k+1}}{4k^2 - 1} T_{2k}(x).
    \end{equation}
\end{lemma}
\begin{proof}
Let $\varepsilon > 0$. Based on \Cref{cor:absvalapprox}, the truncation error $E_d(x) \coloneqq | |x| - f_d(x)| $ can be computed via
   \begin{align*}
       E_d(x) =& \left| |x| - f_d(x) \right| = \left| (\mathcal{F}f)(x) - f_d(x) \right| \\=& \left| \frac{4}{\pi}\sum_{k=d+1}^{\infty} \frac{(-1)^{k+1}}{4k^2 - 1} T_{2k}(x) \right|.
   \end{align*}
Aiming to bound $E_d(x) \leq \varepsilon$, we acknowledge the bound $\left| T_{2k}(x) \right| \leq 1 $ for $x \in [-1, 1] $, such that
\begin{align*}
    E_d(x) \leq & \frac{4}{\pi} \sum_{k=d+1}^{\infty} \left| \frac{(-1)^{k+1}}{4k^2 - 1} \right| \leq \frac{4}{\pi} \sum_{k=N+1}^{\infty} \frac{1}{4k^2 - 1} \\
    = & \frac{4}{\pi} \sum_{k=d+1}^{\infty} \frac{1}{k^2} \frac{1}{4-1/k^2} \leq \frac{4}{3\pi} \sum_{k=d+1}^{\infty} \frac{1}{k^2} \\ \leq & \frac{4}{3\pi} \int_{d}^{\infty} \frac{1}{x^2} dx = \dfrac{4}{3\pi d} \leq \varepsilon
\end{align*}
To satisfy $E_d(x) \leq \varepsilon $, we hence need $d \geq 4 / 3\pi \varepsilon$.
\end{proof}

\end{document}